\documentclass[11pt,a4paper]{iopart}
\usepackage{iopams}
\usepackage[english]{babel}
\usepackage{float}
\usepackage[dvips]{graphicx}
\usepackage{rotating}
\usepackage{epsfig}
\usepackage[dvips]{color}

\newcommand{\lwig}{\mbox{\;\raisebox{.3ex}
    {$<$}$\!\!\!\!\!$\raisebox{-.9ex}{$\sim$}\;}}

\definecolor{Red}{named}{Red}

\begin{document}
\title{Neutrinos in Non-linear Structure Formation - The Effect on Halo Properties}
\author{Jacob Brandbyge$^1$, Steen Hannestad$^1$, Troels Haugb{\o}lle$^2$, Yvonne Y. Y. Wong$^3$}
\ead{jacobb@phys.au.dk, sth@phys.au.dk, haugboel@nbi.dk, yvonne.wong@physik.rwth-aachen.de}
\address{$^1$~Department of Physics and Astronomy, Aarhus University, Ny Munkegade, DK-8000 Aarhus C, Denmark. \\
$^2$~Niels Bohr Institute, Dept. of Astrophysics, Juliane Maries Vej 30, 2100 K\o benhavn  \O, Denmark. \\
$^3$~Institut f\"ur Theoretische Teilchenphysik und Kosmologie, RWTH Aachen, D-52056 Aachen, Germany.}
\date{\today}

%%%%%%%%%%%%%%%%%%%%%%%%%%%%%%%%%%%%%%%%%%%%%%%%%%%%%%%%%%%%%%%%%%%%%%%%%%%%%%%%%%%%%%%%%%%%%%%%%%%%%%%%%%%%%%%%%%%%%%%%%%%%%%%%%%%%%
\begin{abstract}
We use $N$-body simulations to find the effect of neutrino masses on halo properties, and investigate how the density profiles of both the neutrino and the dark matter components change as a function of the neutrino mass. We compare our neutrino density profiles with results from the $N$-one-body method and find good agreement. We also show and explain why the Tremaine-Gunn bound for the neutrinos is not saturated. Finally we study how the halo mass function changes as a function of the neutrino mass and compare our results with the Sheth-Tormen semi-analytic formulae. Our results are important for surveys which aim at probing cosmological parameters using clusters, as well as future experiments aiming at measuring the cosmic neutrino background directly.
\end{abstract}

\section{Introduction}

Massive neutrinos are known to have a significant effect on cosmic structure formation~\cite{Bond:1980ha,Doroshkevich:1980zs}.
In the early universe they contribute to the relativistic energy density and influence the transition from radiation to matter domination. At late times they contribute to the dark matter density, and therefore also to cosmic structure formation. However, as opposed to Cold Dark Matter (CDM), they do not contribute to structure formation on physical scales smaller than the free-streaming scale, roughly equal to the distance traversed before the neutrinos become non-relativistic. This suppression of small-scale structure leaves a very distinct imprint on large-scale structure observables such as the matter power spectrum, which can in turn be used to probe neutrino physics.  Many studies have by now been devoted to this topic, most of which focussing on constraining the neutrino mass, $m_\nu$. At present an upper limit on the neutrino mass can be derived from observations
of the Cosmic Microwave Background (CMB) anisotropies alone or in conjunction with various large-scale structure data sets, such as the SDSS-DR7 LRG catalog, and falls in the range
$\sum m_\nu \lwig 0.4-0.7$ eV, depending both on the complexity of the model space and the combination of data sets used (e.g., \cite{Komatsu:2010fb,Reid:2009nq,Hamann:2010pw}).
In the future the sensitivity of large-scale structure observations to the neutrino mass will increase significantly. For example it has been estimated that the combination of CMB data from Planck and a weak lensing survey from the LSST will push the 1$\sigma$ sensitivity to better than 0.05~eV, close to the minimum $\sum m_\nu$ allowed by oscillation data~\cite{Hannestad:2006as}.

While most neutrino mass constraints at present have been derived
using large-scale structure correlation functions (or power spectra), there are other observables that are potentially just as interesting. One prime example is cluster number counts which are in principle very sensitive to the neutrino mass \cite{Wang:2005vr}. However, in order to fully utilise such data it is necessary to have accurate theoretical predictions, which so far do not exist for $\Lambda$CDM models extended with massive neutrinos (see, however, \cite{Kofman:1995ds} for an early calculation based on the old mixed dark matter scenario). In the present paper we calculate the halo mass function in $\Lambda$CDM cosmologies with massive neutrinos included for a variety of neutrino masses.
However, before proceeding to this and a discussion of other observables related to halo properties, let us briefly review how neutrinos affect structure formation in the linear regime.

 \subsection{The effect of neutrinos}
The effect of neutrinos on structure formation in linear theory has been studied numerous times in the literature (see, e.g., \cite{Lesgourgues:2006nd}).
In general the amount of fluctuations at a given wavenumber $k$ is represented by the power spectrum, $P(k) = |\delta_k|^2$, which can be split in the following form
\begin{equation}
P(k,z)=D(z) T^2(k,z) P_0(k),
\end{equation}
where $D(z)$ is a scale independent growth factor and $P_0$ is the initial power spectrum. $T(k)$ is the transfer function (TF) which is both time and scale dependent in general. The effect of massive neutrinos is embedded entirely in the TF and is separated into two regimes. On scales much larger than the free-streaming scale,
\begin{equation}
k_{\rm FS} \sim 0.8 \frac{m_\nu}{\rm eV} \, h \, {\rm Mpc}^{-1},
\end{equation}
where $m_\nu$ is the one-particle neutrino mass, neutrinos behave essentially like CDM, while on smaller scales they suppress structure formation.
Very na\"{\i}vely one might expect the suppression arising from replacing a fraction of the CDM component with neutrinos to be of order $\Delta P/P \sim -\Omega_\nu/\Omega_m$,
where $\Omega_{m} = \Omega_c+\Omega_b+\Omega_\nu$ is the total matter density,
because neutrinos do not cluster. However, this grossly underestimates the true effect because massive neutrinos also influence the background expansion around the time of matter-radiation equality. The final result in linear theory is that the suppression is approximately given by
\begin{equation}
\frac{\Delta P}{P} \sim - 8 \frac{\Omega_\nu}{\Omega_m}.
\end{equation}
This shows that most of the effect actually comes from the modification to the background, i.e., sub-eV to eV scale neutrinos lead to a longer radiation era.
This effect is also much larger than the effect of  replacing a fraction $\Delta \Omega_m$ of the CDM energy density with $\Lambda$. In this case the change in the matter power spectrum on small scales
(with the large-scale normalisation held constant, i.e., ignoring the effects of $\Delta \Omega_m$ on the growth factor) is very approximately given by
\begin{equation}
\frac{\Delta P}{P} \sim \left(\frac{\Omega_{m}^{'}}{\Omega_{m}}\right)^{7/2} \sim - 3.5 (1-\Omega_{m}^{'}/\Omega_{m}),
\label{eq:cdm}
\end{equation}
where $\Omega_{m}^{'} = \Omega_{m}+\Delta \Omega_m$, for small changes in $\Delta \Omega_m \ll \Omega_m$. The effect here is approximately two times smaller than that due to assigning $\Delta \Omega_m$ to massive neutrinos.

Since neutrinos have such a strong effect on the power spectrum even in linear theory it is natural to expect a similarly strong effect in the non-linear regime. This was tested in detail for the power spectrum in a number of papers \cite{Brandbyge1,Brandbyge2,Brandbyge3}, and a significant enhancement in the power spectrum suppression was indeed found: The maximum suppression is increased from $-8 \Omega_\nu/\Omega_{m}$ to approximately $-9.8 \Omega_\nu/\Omega_{m}$,\footnote{This finding was confirmed in the very recent paper \cite{Viel:2010bn}.} with a pronounced feature at $k \sim 0.7 \, h \, {\rm Mpc}^{-1}$.

Another issue which has so far not been addressed with precision $N$-body simulations is how the presence of massive neutrinos affect halo formation. Here we study how CDM halo properties are altered by the presence of massive neutrinos, and we also present detailed results for the corresponding neutrino halos. This last point is important for example for understanding the prospects for a direct experimental detection of the cosmic relic neutrino background.

The paper is organised as follows: In Section \ref{sec:numericalsetup} we present the numerical setup required for the analysis. In Section \ref{sec:halostructure} we present results on halo profiles for both the neutrino and matter components. In Section \ref{sec:massfunction} we discuss how the halo mass function is altered in models with massive neutrinos, and finally Section \ref{sec:conclusion} contains our conclusions.

\section{Numerical setup}
\label{sec:numericalsetup}

\subsection{Initial conditions and $N$-body simulations}

Our $N$-body simulations are carried out using the modified version of the \textsc{gadget}-2 code \cite{Springel:2000yr} described in \cite{Brandbyge3}. The code utilises a hybrid scheme for simulating the neutrino component: All neutrinos with velocities much higher than the average gravitational flow velocities in the simulation are treated using linear perturbation theory, and their effect on the gravitational potential in the $N$-body simulation is included via a Fourier grid. At low redshift the low velocity neutrinos, with $q/T < 6$, are followed separately in 6 momentum bins by $N$-body particles ($q$ is  the comoving momentum and $T$ the comoving temperature). In Fig.~\ref{fig:nuprofile2} it can be seen that the higher part of momentum space, $q/T > 6$, do not contribute to neutrino clustering in halos.

The initial conditions (ICs) for the CDM and neutrino components are generated with the same set of random numbers for a given box size. This reflects the assumption of adiabatic primordial ICs. The CDM distribution is followed with CAMB \cite{CAMB} until $z = 49$, where the Zel'dovich approximation \cite{Zeldovich:1969sb} and a second-order correction calculated with second-order Lagrangian perturbation theory \cite{Scoccimarro1} are used to generate the CDM $N$-body particle initial displacements and gravitational flow velocities.

When the maximum thermal velocity of the first neutrino momentum bin, $q/T = 1$ (upper limit), has fallen below $f_{\rm flow} = 4$ times the average CDM gravitational velocity,\footnote{When the box size is increased the average CDM gravitational velocity increases as well due to extra large-scale velocity flows. As a result, the neutrino $N$-body particles are created earlier in large simulation volumes, and not due to larger non-linearities which of course would justify a larger conversion redshift. But this effect is small, at the order of 5\% in redshift. Ideally, the velocities should be interpolated to a grid, Fourier transformed and then convolved with a window function eliminating large-scale velocity modes.} the low momentum neutrinos, which have non-linear clustering, are converted to $N$-body particles with the Zel'dovich approximation and followed with \textsc{gadget}-2. The high velocity part is retained on the grid. The ICs for each bin are generated from a momentum dependent TF, and in addition to the gravitational flow velocities the neutrino $N$-body particles receive a thermal, Fermi-Dirac distributed, velocity corresponding to the particular bin. In effect we are simulating 6 different Hot Dark Matter components with different thermal properties. We employ a new timestep criterion for the neutrino component $\propto (1+z) / \Omega_\nu$.

The TF for each momentum bin is calculated with CAMB. By default CAMB uses 15 neutrino momentum bins, which is sufficient for sub-percent accuracy in the momentum {\it averaged} TFs. However, for our purpose, we need percent level accuracy in each individual momentum bin, and to ensure this we use a total of 480 individual TFs such that each of the 15 momentum bins is reconstructed from 32 separate TFs. See \cite{Brandbyge3} for further information on the hybrid implementation of neutrinos in $N$-body simulations.

The hybrid method has the advantage that all neutrino masses can be simulated without compromising computational speed or accuracy (treating low mass neutrinos as particles leads to prohibitively high CPU time consumption\footnote{Further investigations for an appropriate Courant timestep condition for the neutrino component could alleviate this problem.} and particle shot noise, and, conversely, treating high mass neutrinos in linear perturbation theory leads to loss of accuracy).
The code allows us to accurately calculate halo number densities and properties over a wide mass range. Since neutrinos cluster much less than CDM ultrahigh resolution is in most cases not necessary.\footnote{Ultrahigh resolution can be achieved by making zoom simulations, though for neutrinos the advantage of such a setup is smaller, since fast-moving neutrinos can easily leak out of the high-resolution region, and hence the initial volume of the zoomed neutrinos has to be considerably larger than the initial CDM volume (though a periodic volume around the halos could alleviate this problem at higher redshift).} For example, since the neutrino halo profiles flatten at small $r$ one does not encounter the same resolution problem as with CDM. Likewise there is hardly any neutrino halo substructure because neutrino halos are primarily formed by late time neutrino infall on already existing CDM structures.

%%%%%%%%%%%% new table %%%%%%%%%%%
\begin{table*}[t]
{\footnotesize
  \hspace*{0.0cm}\begin{tabular}
  {c||c@{\hspace{11pt}}c@{\hspace{11pt}}c@{\hspace{11pt}}c@{\hspace{11pt}}c@{\hspace{11pt}}c@{\hspace{11pt}}c@{\hspace{11pt}}c@{\hspace{11pt}}c} \hline
%%%%%%%%%%%%%%%%%
  & $N_{\rm CDM}$ & $N_{\nu,{\rm grid}}$ & $N_{\nu,{\rm part}}$ & $q_{\rm cut}/T$ & $f_{\rm flow}$ & $R_{\rm BOX}$ $[h^{-1} {\rm Mpc}]$ & $\sum m_\nu~[{\rm eV}]$ & $\Omega_{\nu}$~[$\%$] & $\Omega_{m}$~[$\%$] \\       \hline\hline
%%%%%%%%%%%%%%%%%
$A_1$   & $512^3$ & 0       & 0              & 0 & - & $ 256$ & 0    & 0     & 30    \\
$A_2$   & $512^3$ & $512^3$ & 0              & 0 & 0 & $ 256$ & 0.15 & 0.325 & 30    \\
$A_3$   & $512^3$ & $512^3$ & 0              & 0 & 0 & $ 256$ & 0.3  & 0.65  & 30    \\
$A_4$   & $512^3$ & $512^3$ & $6\cdot 512^3$ & 6 & 4 & $ 256$ & 0.6  & 1.3   & 30    \\
$A_5$   & $512^3$ & $512^3$ & $6\cdot 512^3$ & 6 & 4 & $ 256$ & 1.2  & 2.6   & 30    \\\hline

$B_1$   & $512^3$ & 0       & 0              & 0 & - & $1024$ & 0    & 0     & 30    \\
$B_2$   & $512^3$ & $512^3$ & 0              & 0 & 0 & $1024$ & 0.15 & 0.325 & 30    \\
$B_3$   & $512^3$ & $512^3$ & 0              & 0 & 0 & $1024$ & 0.3  & 0.65  & 30    \\
$B_3^{'}$&$512^3$ & $512^3$ & $6\cdot 512^3$ & 6 & 4 & $1024$ & 0.3  & 0.65  & 30    \\
$B_4$   & $512^3$ & $512^3$ & $6\cdot 512^3$ & 6 & 4 & $1024$ & 0.6  & 1.3   & 30    \\
$B_5$   & $512^3$ & $512^3$ & $6\cdot 512^3$ & 6 & 4 & $1024$ & 1.2  & 2.6   & 30    \\\hline

$C_1$   & $512^3$ & 0       & 0              & 0 & - & $4096$ & 0    & 0     & 30    \\
$C_2$   & $512^3$ & $512^3$ & 0              & 0 & 0 & $4096$ & 0.15 & 0.325 & 30    \\
$C_3$   & $512^3$ & $512^3$ & 0              & 0 & 0 & $4096$ & 0.3  & 0.65  & 30    \\
$C_4$   & $512^3$ & $512^3$ & $6\cdot 512^3$ & 6 & 4 & $4096$ & 0.6  & 1.3   & 30    \\
$C_5$   & $512^3$ & $512^3$ & $6\cdot 512^3$ & 6 & 4 & $4096$ & 1.2  & 2.6   & 30    \\\hline

$D_1$   & $512^3$ & 0       & 0              & 0 & - & $ 256$ & 0    & 0     & 28.7  \\
$D_2$   & $512^3$ & 0       & 0              & 0 & - & $1024$ & 0    & 0     & 28.7  \\
$D_3$   & $512^3$ & 0       & 0              & 0 & - & $4096$ & 0    & 0     & 28.7  \\\hline

$E_1$   & $512^3$ & 0       & 0              & 0 & - & $ 256$ & 0.6  & 1.3   & 30    \\
$E_2$   & $512^3$ & 0       & 0              & 0 & - & $1024$ & 0.6  & 1.3   & 30    \\
$E_3$   & $512^3$ & 0       & 0              & 0 & - & $4096$ & 0.6  & 1.3   & 30    \\
$E_4$   & $512^3$ & 0       & 0              & 0 & - & $ 256$ & 0.6  & 1.3   & 30    \\
$E_5$   & $512^3$ & 0       & 0              & 0 & - & $1024$ & 0.6  & 1.3   & 30    \\
$E_6$   & $512^3$ & 0       & 0              & 0 & - & $4096$ & 0.6  & 1.3   & 30    \\\hline
  \end{tabular}
  }
  \caption{$N$-body simulation parameters: $N_{\rm CDM}$ and $N_{\nu,{\rm part}}$ are the number of CDM and neutrino $N$-body particles respectively, $N_{\nu, {\rm grid}}$ the size of the linear neutrino Fourier grid, $q_{\rm cut}/T$ the cut-off below which the neutrino component is converted to particles, and $f_{\rm flow}$ determines the redshift of this conversion. The simulation box size is represented by $R_{\rm BOX}$, $\sum m_\nu$ is the total neutrino mass roughly related to the neutrino density parameter, $\Omega_\nu$, by $\Omega_\nu = \sum m_\nu / (94 \, h^2 {\rm eV})$, with $\Omega_m = \Omega_c + \Omega_b + \Omega_\nu$. The exotic simulations $E_{1-3}$ have no $\delta_\nu$ in the $N$-body simulation, but the neutrinos are still included in the background evolution, while $E_{4-6}$ have a CDM $N$-body particle mass corresponding to $\Omega_{m} = 0.3$ and the ICs are calculated from weighed CDM and baryon TFs from a cosmology with neutrinos included.}
  \label{fig:table1}
\end{table*}

\paragraph{Parameter setup}

We assume a standard flat cosmology with $h = 0.7$, $\sigma_8=0.878$ (for a model without massive neutrinos), $n_{\rm s}=1$, $A_{\rm s}=2.3\cdot 10^{-9}(\Omega_{m}/0.3)^2$, $\Omega_{b} = 0.05$ and varying amounts of $\Omega_{c}$ and $\Omega_\nu$. In cosmologies with massive neutrinos we assume 3 degenerate neutrino species. Table \ref{fig:table1} shows parameters for the various $N$-body simulations presented in this paper.

By running simulations in simulation volumes of 32, 128 and $256 \, h^{-1} \, {\rm Mpc}$ we found that the neutrino halo profiles were almost identical in the latter two box sizes, and that the smallest box significantly affected the density profiles for the larger halos. In sum, we only present results for simulation volumes larger than $256 \, h^{-1} \, {\rm Mpc}$.

We have chosen $f_{\rm flow} = 4$ as the criterion for creating neutrino $N$-body particles, which we have shown to be a reasonable value by comparing with simulations in which $f_{\rm flow} = 2$ and 8.

Finally, when we present matter density profiles for cosmologies with $\sum m_\nu = 0.15 \, {\rm eV}$ and $0.3 \, {\rm eV}$ neutrinos, only the homogeneous neutrino component has been added. This seems reasonable since these low mass particles contribute insignificantly to the overall matter density profile.

\begin{figure}
    \includegraphics[width=0.45\linewidth, angle = 90]{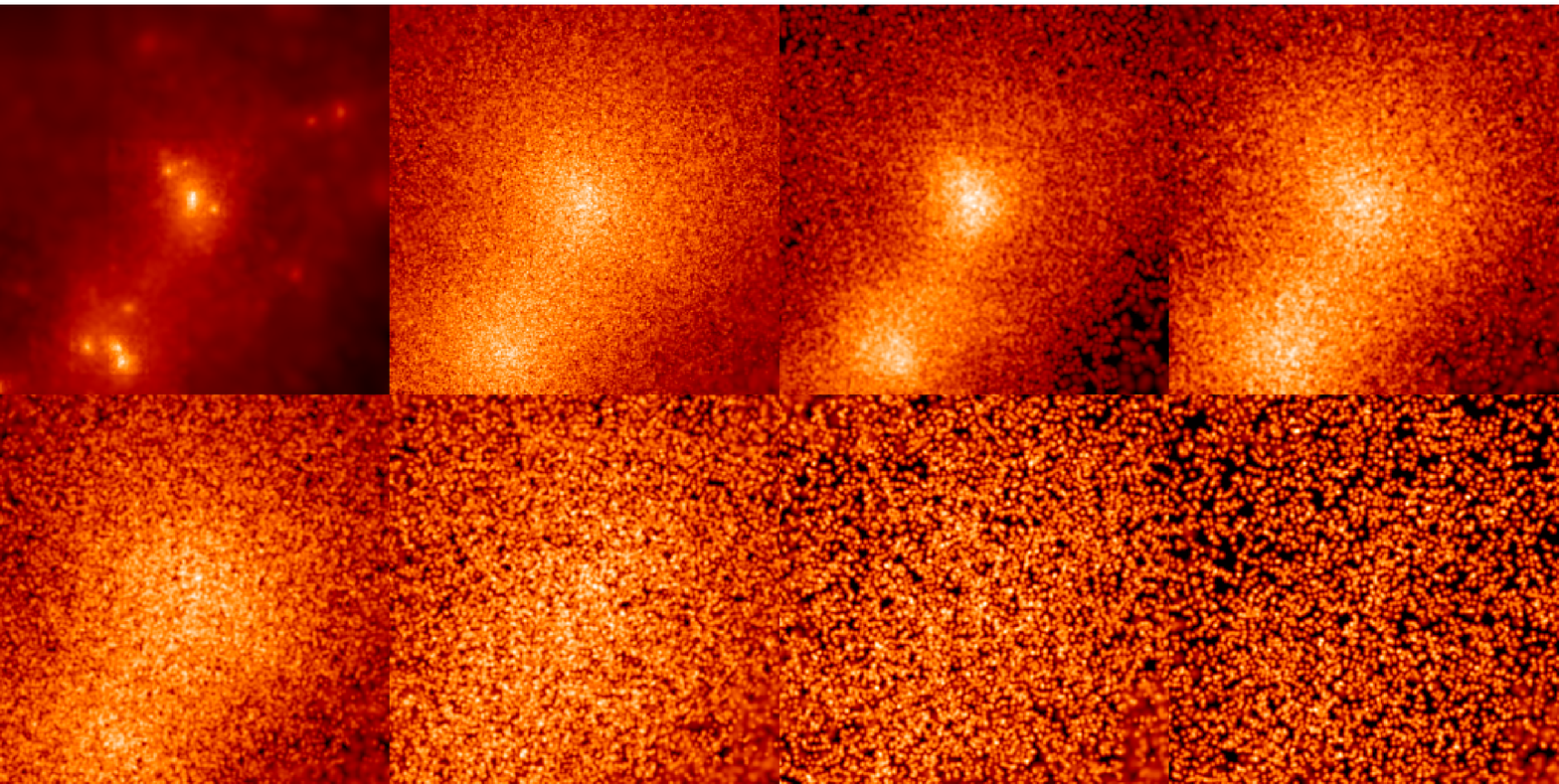}\vspace*{0.1cm}
    \includegraphics[width=0.45\linewidth, angle = 90]{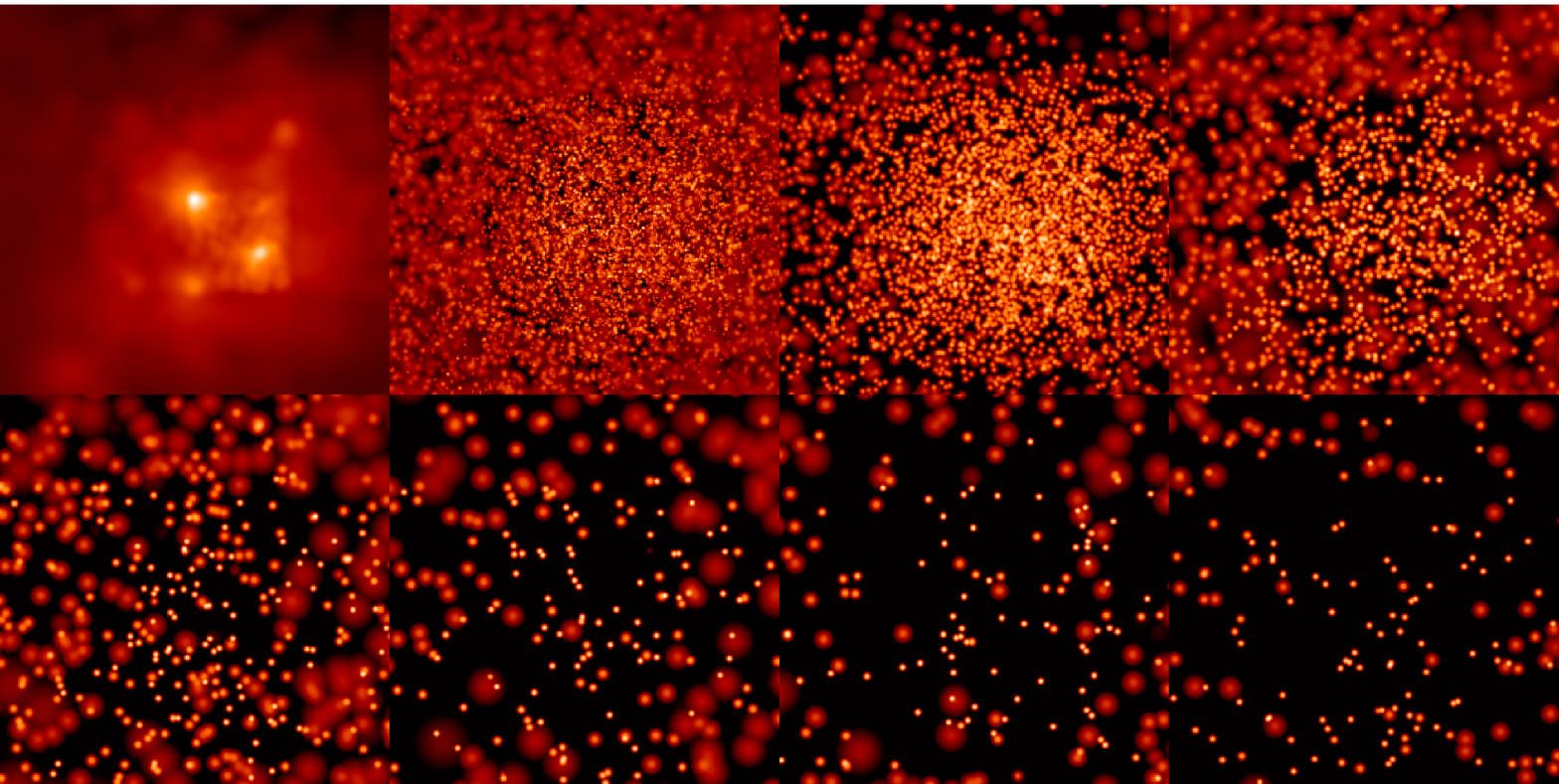}\vspace*{0.1cm}
    \includegraphics[width=0.45\linewidth, angle = 90]{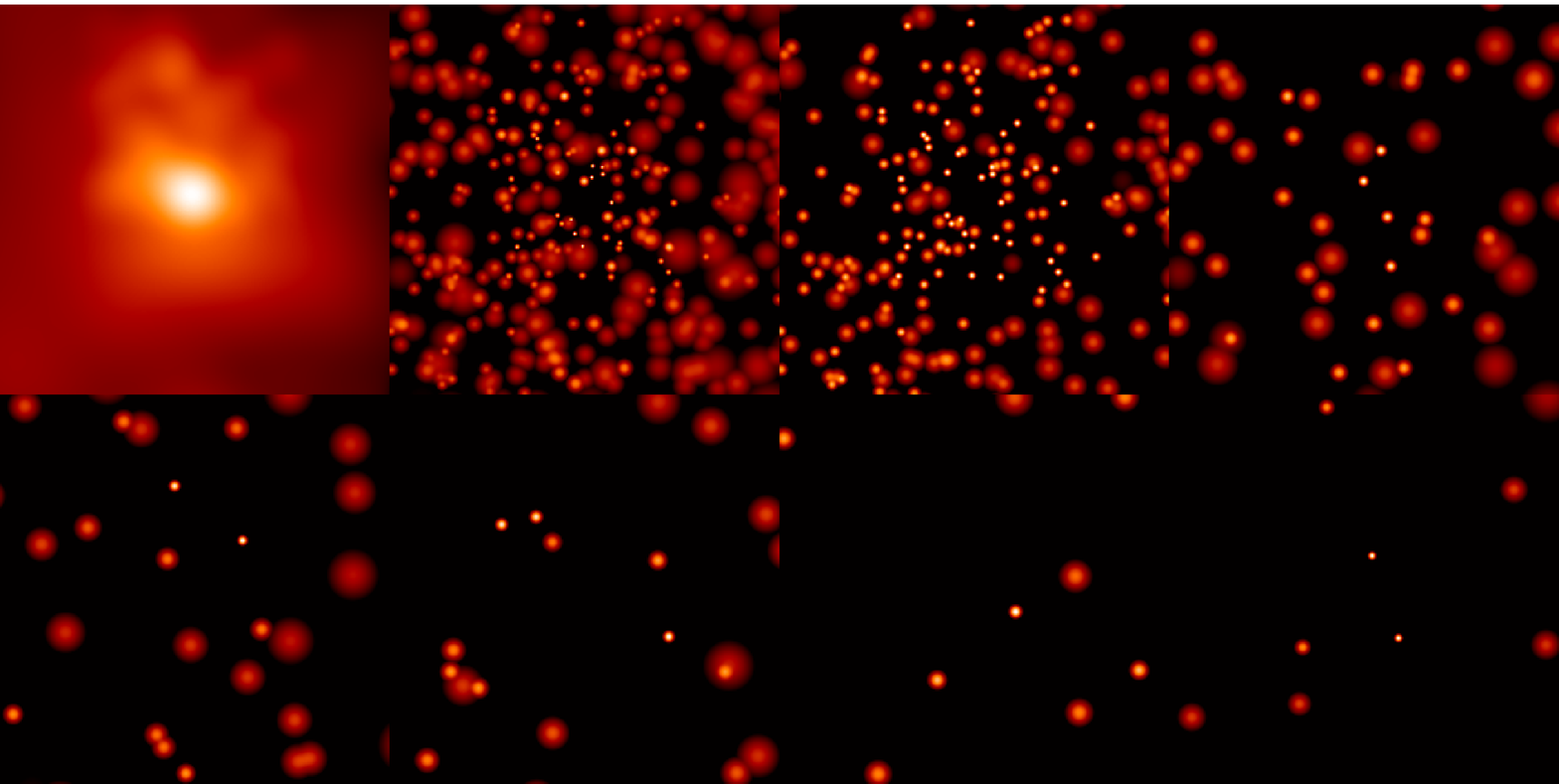}
    \caption{CDM and $\sum m_\nu = 1.2\, {\rm eV}$ neutrino distributions for halo masses $\simeq 5\cdot 10^{14}{\rm M}_\odot$ (top), $\simeq 10^{14}{\rm M}_\odot$ (middle) and $\simeq 10^{13}{\rm M}_\odot$ (bottom), where the masses only correspond to the central halos in the upper two mosaics. Dimensions in each image are 5, 2 and 1 $h^{-1} \, {\rm Mpc}$, respectively. In each mosaic the images correspond to CDM, total neutrino, and $q/T = 1$ to 6 from top-left to bottom-right. Individual neutrino $N$-body particles can be identified.}
   \label{fig:image1}
\end{figure}

Fig.~\ref{fig:image1} shows the CDM and neutrino $N$-body particles in halos with different masses. Here, individual $N$-body particles can be identified, with small and bright particles lying in high density regions, and larger and darker particles in lower density areas.

Fig.~\ref{fig:image1} clearly illustrates the lack of neutrino particle statistics in halo centers. Neutrino clustering is determined by the combined effects of gravity and thermal velocity. The smaller the neutrino and the halo masses the less clustering. To simulate such small overdensities a very fine initial $N$-body particle grid is required. From the figure it can also be seen that only $q/T \lesssim 3$ trace the underlying CDM distribution on the scales shown, whereas structure in the higher momentum bins can only be seen on larger scales. This is also consistent with Fig.~\ref{fig:nuprofile2}.

\subsection{Halo finding}
We use the A\textsc{miga} Halo Finder (AHF) \cite{Knollmann:2009pb} to identify halos and their centers within the simulations. Only bound particles have been used to calculate halo centers and their virial radii. From the halo centers we calculate the matter and the neutrino halo profiles for halo masses of $10^{12}$,  $10^{13}$, $10^{14}$ and $10^{15} \, {\rm M}_\odot$. We show matter halos and not CDM halos, since it is the former quantity which is the measurable one.

We stack halos in mass bins with a bin width of 10\% of the central halo mass. These widths are narrow enough to ensure a reliable halo profile calculation of the desired halo masses, while at the same time providing enough halos to stack. The $N$-body density profiles presented in this paper have been found by fitting a smooth curve through the data points, though we note that this is mainly important in the inner part of the neutrino density profiles for low neutrino and halo masses, where particle statistics is low.

Furthermore, to compare with the $N$-one-body method, we have only used halos with an average overdensity within the virial radius, $\Delta_{\rm vir}$, in the range $330-340$ for $\Omega_m = 0.3$ ($340-350$ for $\Omega_m = 0.287$). This criterion eliminates, e.g., gravitationally stripped halos with a very high central density.

\subsection{The $N$-one-body method}

The $N$-one-body method was introduced in~\cite{Ringwald:2004np}.  It is a restricted method
devised to solve, approximately, the (non-linear) collisionless Boltzmann equation for the neutrino phase space based on the
following observation:
In the limit $\rho_\nu \ll \rho_m$ density perturbations in the CDM fluid dominate the
total gravitational potential, and not only will the CDM halo be gravitationally blind to the
neutrinos, gravitational interaction between the neutrinos themselves will also be negligible.
This allows us to treat the CDM halo as an external gravitational source, and
compute the trajectory of each neutrino phase space element as a test particle moving in an
external potential one at a time in $N$ independent simulations.
An obvious advantage of this technique is that it requires virtually no computing power when
compared with a full scale $N$-body simulation with the same, large $N$, and is thus particularly
useful for resolving the clustering of neutrinos on small scales.

In principle the $N$-one-body method can be applied to any given CDM density distribution.  Here, however,
we focus exclusively on spherically symmetric CDM halos whose density profiles are parameterised by
fitting functions.  Apart from possible deviations in the innermost cores, dark matter halos are well
described by the universal Navarro-Frenk-White (NFW) profile~\cite{Navarro:1996gj}, with a density given by
\begin{equation}
	\rho_{\rm halo}(r)=\frac{\rho_{\rm s}}{(r/r_{\rm s})(1+r/r_{\rm s})^2},
	\label{eq:NFW}
\end{equation}
where $\rho_{\rm s} = 4\rho(r_{\rm s})$ is a characteristic inner density at the characteristic radius $r_{\rm s}$. $r_{\rm s}$ is related to the virial radius $r_{\rm vir}$ via the concentration parameter $c$
\begin{equation}
	c\equiv\frac{r_{\rm vir}}{r_{\rm s}}.
	\label{eq:c}
\end{equation}
In a $\Lambda$CDM cosmology, a good analytical expression for the concentration parameter is~\cite{Bullock:1999he}
\begin{equation}
	c(z=0)\simeq9\biggl(\frac{M_{\rm vir}}{1.5\times 10^{13} h^{-1}{\rm M}_\odot}\biggr)^{-0.13},
	\label{eq:cfit}
\end{equation}
where $M_{\rm vir}$ is the virial mass lying within $r_{\rm vir}$, and
$c(z) \sim c(0)/(1+z)$.  We expect $c$ to be smaller for cosmologies with neutrinos since the neutrinos free-stream out of the inner density cores and effect the halo formation process.
Defining the average overdensity within $r_{\rm vir}$ to be $\Delta_{\rm vir}$, we have
\begin{equation}
	M_{\rm vir} \equiv  \frac{4\pi}{3}\Delta_{\rm vir}\bar{\rho}_{{\rm m},0}r_{\rm vir}^3 = 4\pi\rho_{\rm s}a^3 r_{\rm s}^3 \biggl[{\rm ln}(1+c)-\frac{c}{1+c}\biggr],
	\label{eq:Mvir}
\end{equation}
so that a halo of a given mass is fully described in terms of its concentration parameter alone.
The overdensity $\Delta_{\rm vir}$ can also be approximated by the overdensity at virialisation from the spherical top-hat collapse model,
\begin{equation}
	\delta_{\rm th} \simeq \frac{18\pi^2+82[\Omega_{m}(z)-1]-39[\Omega_{m}(z)-1]^2}{\Omega_{m}(z)},
	\label{eq:tophat}
\end{equation}
with $\Omega_m(z) = \Omega_m/(\Omega_m + \Omega_\Lambda a^3)$~\cite{Bryan:1997dn}.
For $\Omega_{m}=0.3$, we find $\delta_{\rm th}\simeq337$ ($\delta_{\rm th}\simeq346$ for $\Omega_{m}=0.287$), which explains why we only use the range $\Delta_{\rm vir} = 330-340$ ($340-350$) when analysing $N$-body data.

We model the CDM distribution as a NFW halo sitting on top of a uniform
distribution of CDM, i.e., the $N$-one-body method assumes that all halos are completely isolated.
In order that the halo overdensity merges smoothly into the background density, we
extend the NFW profile to beyond the virial radius.   The initial neutrino distribution is
taken to be spatially uniform,\footnote{We have run one $N$-body simulation where the neutrino particles were assigned homogeneous ICs, which in general confirmed the validity of this assumption.} with a momentum distribution described by relativistic
Fermi-Dirac statistics.  This initial distribution is divided into small chunks in both real and momentum space,
and each chunk is allowed to move under the external potential of the CDM halo, but independently of
each other. A low resolution run is first carried out for each set of neutrino and halo masses.
All chunks that end up at $z = 0$ inside a sphere of radius $50 \ h^{-1} \ {\rm Mpc}$ centered on the halo
are traced back to their origin, subdivided into smaller chunks, and then re-simulated. The process is repeated
until the inner $\sim 10 \ h^{-1} \, {\rm kpc}$ is resolved.

The $N$-one-body simulations are started at the same redshifts as when the neutrino $N$-body particles are created. This redshift is neutrino mass and halo mass dependent. Note that when comparing the density profiles with the two methods, the underlying CDM profiles are not identical, since the neutrino component do contribute to the gravitational potential and not least the halo merger history in the $N$-body simulations.

%%%%%%%%%%%%%%%%%%%%%%%%%%%%%%%%%%%%%%%%%%%%%%%%%%%%%%%%%%%%%%%%%%%%%%%%%%%%%%%%%%%%%%%%%%%%%%%%%%%%%%%%%%%%%%%%%%%%%%%%%%%%%%%
\section{Halo structure}
\label{sec:halostructure}

\begin{figure}
     \noindent
     \vspace*{-1cm}
     \begin{center}
           \includegraphics[width=0.7\linewidth]{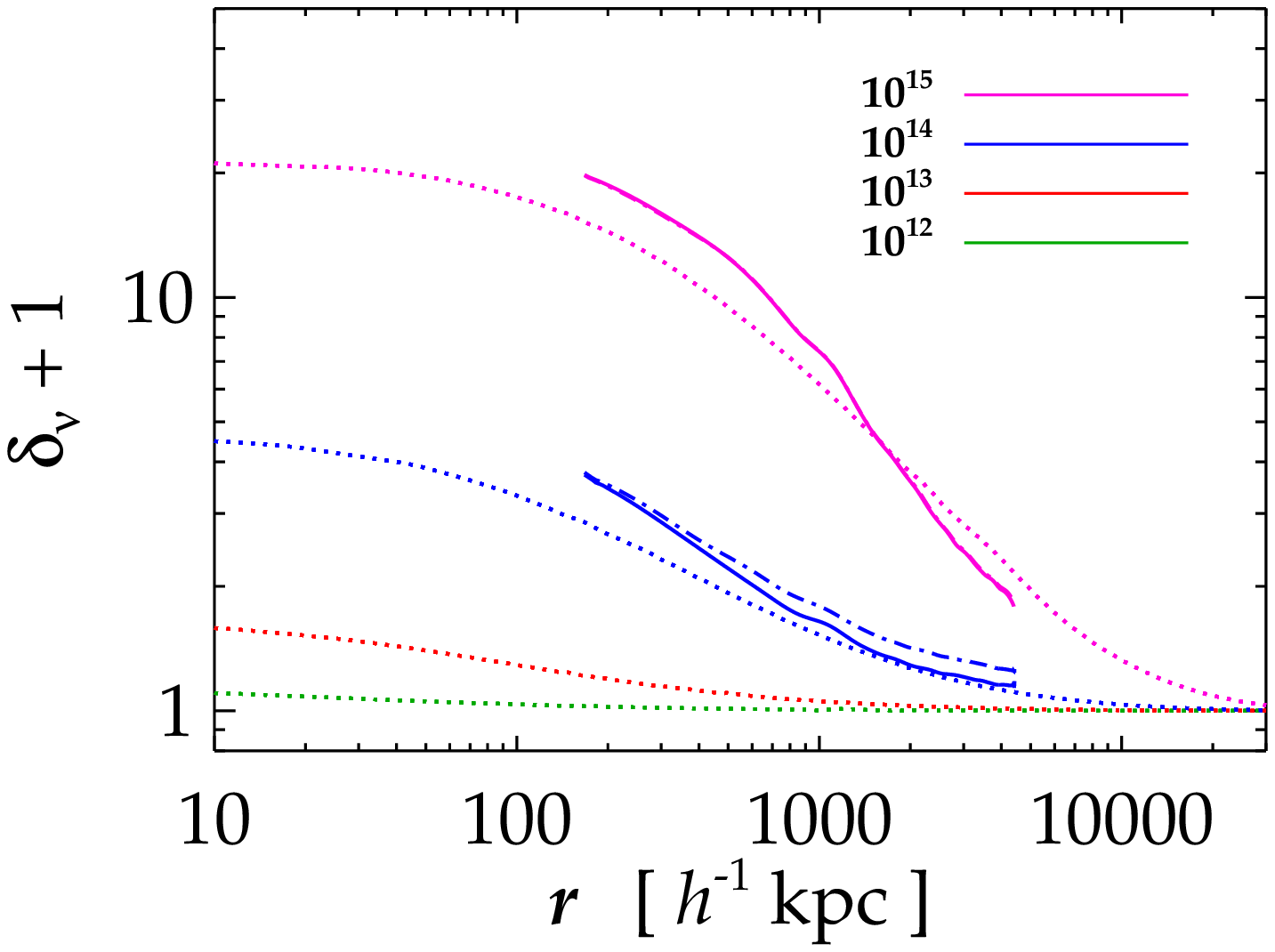}
     \end{center}
     \vspace*{-1cm}

     \begin{center}
           \includegraphics[width=0.7\linewidth]{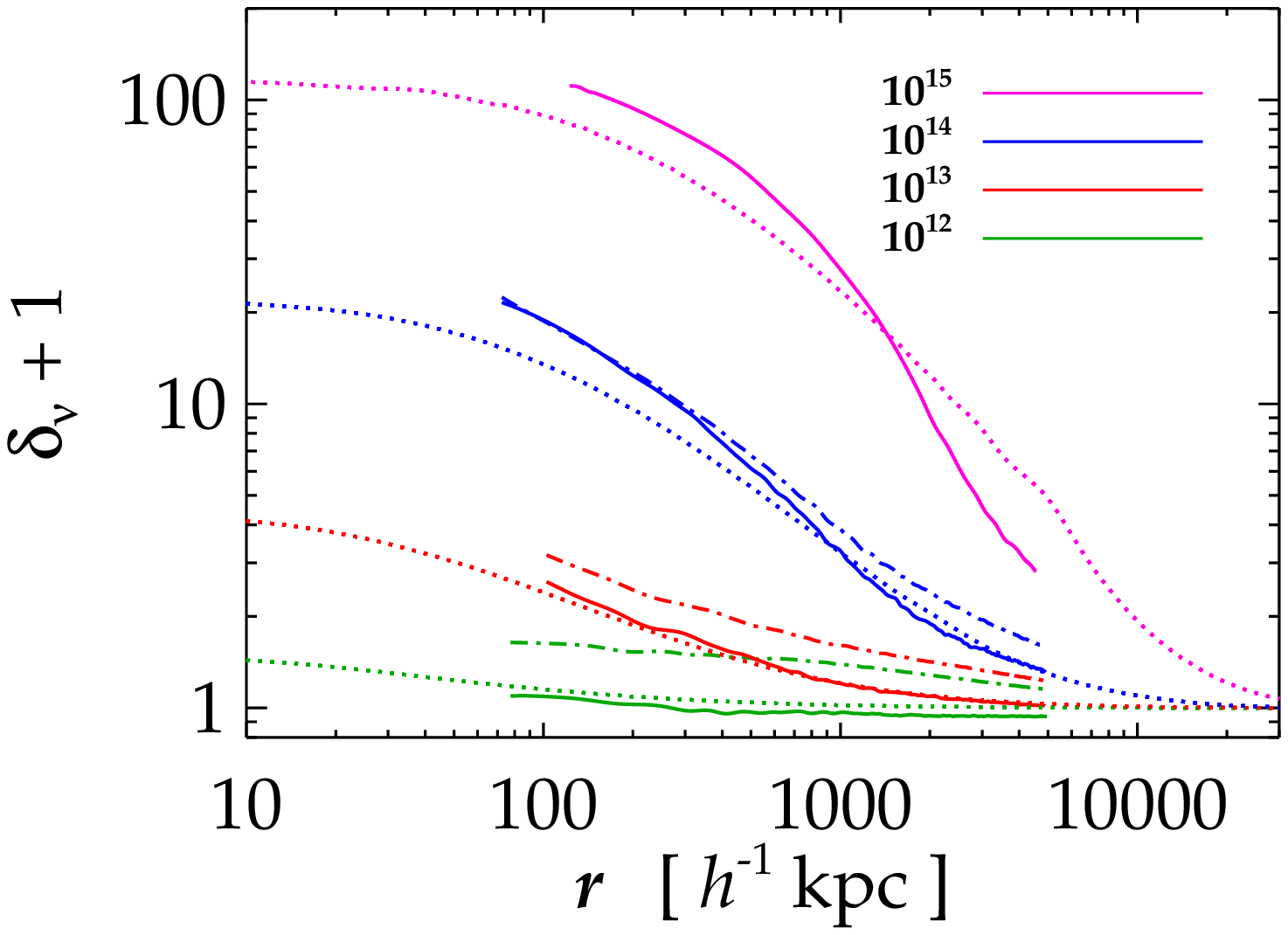}
     \end{center}
     \vspace*{-1cm}

     \begin{center}
           \includegraphics[width=0.7\linewidth]{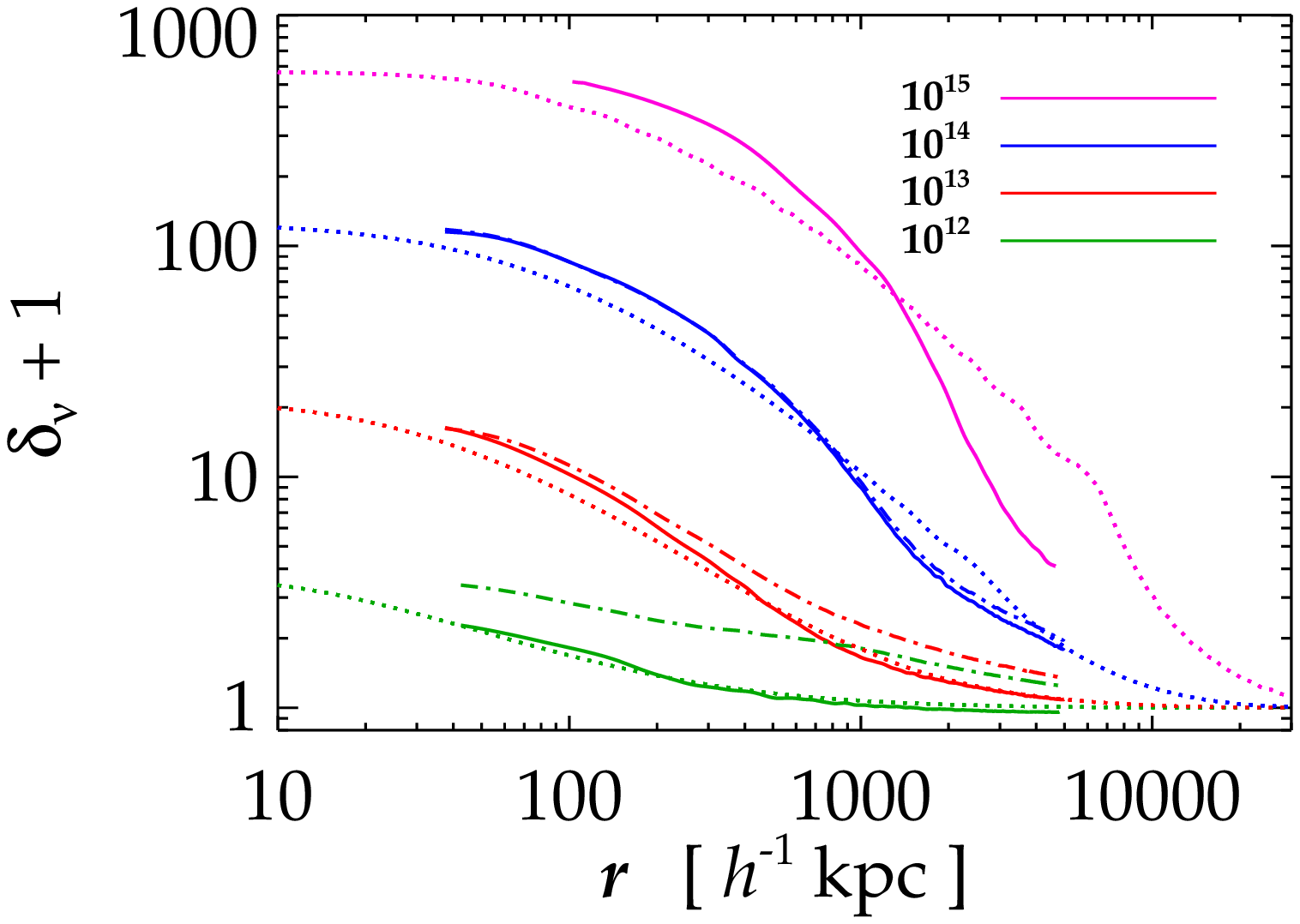}
     \end{center}
     \vspace*{-0.5cm}
     \caption{Neutrino halo profiles for $\sum m_\nu = 0.3 \, {\rm eV}$ (top),  $\sum m_\nu = 0.6 \, {\rm eV}$ (middle) and $\sum m_\nu = 1.2 \, {\rm eV}$ (bottom) for halo masses of $10^{12}$,  $10^{13}$, $10^{14}$ and $10^{15} \, {\rm M}_\odot$. Profiles are calculated with the $N$-one-body method (dotted) and the $N$-body method with a halo isolation criterion (solid) and without (dot-dashed).}
   \label{fig:nu_profiles}
\end{figure}

\begin{figure}
    \noindent
    \begin{center}
           \includegraphics[width=0.8\linewidth]{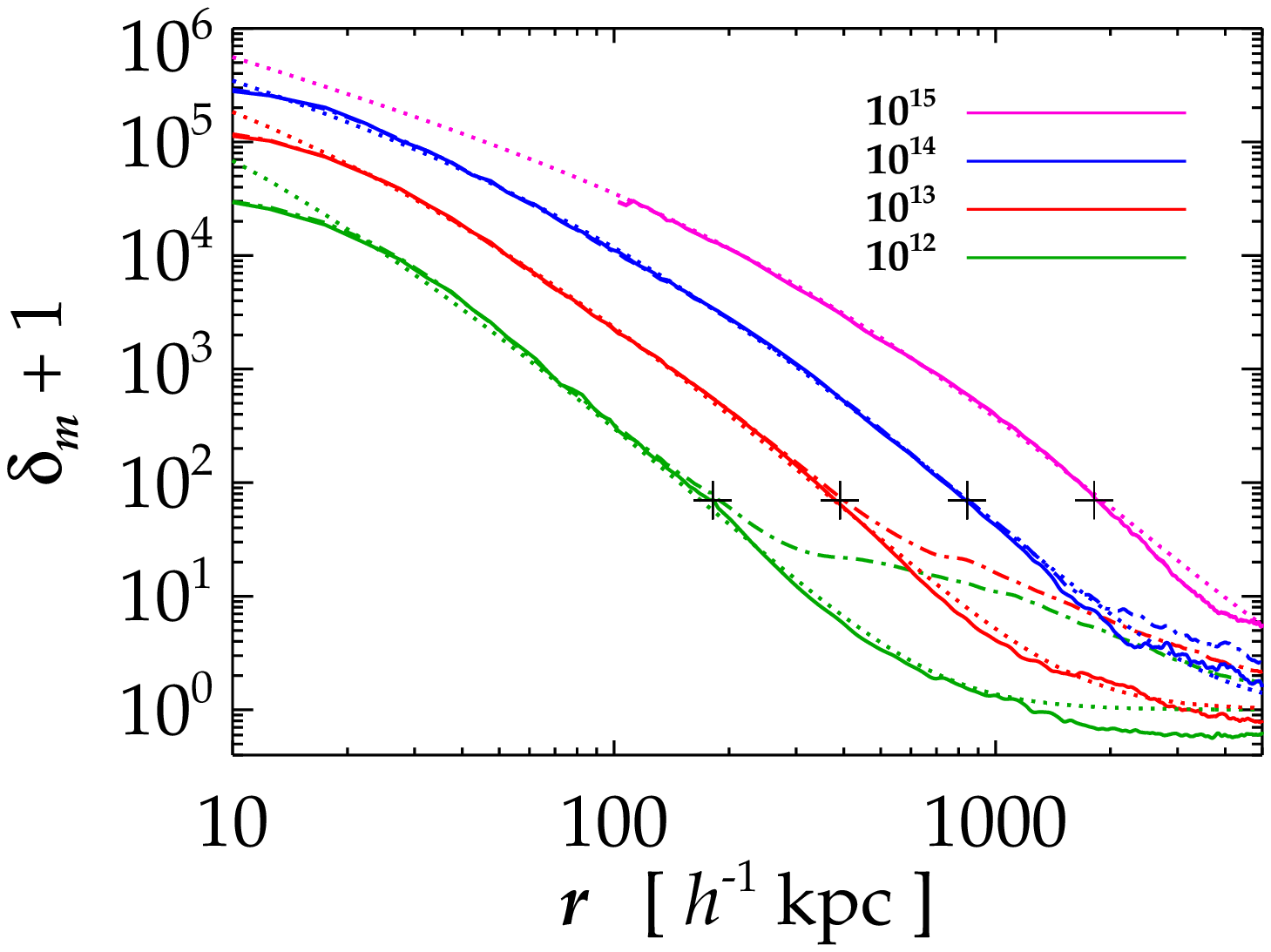}
    \end{center}
   \caption{Halo profiles from $N$-body simulations for a model without massive neutrinos, with isolated halos (solid) and all halos (dot-dashed). The halo masses are $10^{12}$,  $10^{13}$, $10^{14}$ and $10^{15} \, {\rm M}_\odot$. The profiles for the lowest 3 halo masses are taken from the $256 \, h^{-1} \, {\rm Mpc}$ box and the profile for the most massive halo is taken from the $1024 \, h^{-1} \, {\rm Mpc}$ box. NFW profiles are also shown (dotted), and the halo mass dependent virial radii are indicated by the '+' signs.}
   \label{fig:NFW}
\end{figure}

\subsection{Neutrino clustering}

The gravitational effect of a host halo is relatively much more important for neutrinos than for the CDM component: Due to free-streaming neutrinos will almost completely stream out of small halos ($\simeq 10^{12}{\rm M}_\odot$), and any measured value $\delta_\nu > 0$ will be caused by the host halo. The radial profile of $\delta_\nu$ will therefore be a superposition of a dominant flat profile from the host halo on top of a sub-dominant contribution from the $\simeq 10^{12}{\rm M}_\odot$ halo itself. This fact can be seen in Fig.~\ref{fig:nu_profiles}.

Since it is the isolated halo profile that the $N$-one-body method calculates, we have found the neutrino density profiles from isolated halos in the $N$-body simulations. We used the criterion that for a halo to be considered isolated it should be more than 10 times the virial radius away from a heavier halo. With this criterion the agreement between the two methods improves $significantly$ for masses $\lesssim 10^{13}{\rm M}_\odot$, in effect confirming the robustness of both the $N$-body and the $N$-one-body methods. It can also be seen that for $\sum m_\nu = 0.3 \, {\rm eV}$ there is a significant contribution to $\delta_\nu$ even in a $\simeq 10^{14}{\rm M}_\odot$ halo from heavier halos, which is not the case for more massive neutrino states. This is caused by the fact that as the neutrino mass decreases, neutrinos free-stream out of ever larger halos, so that the relative effect of even larger host halos must be taken into account.

The effect of tidal truncation on the $N$-body halo profiles can easily be seen in Fig.~\ref{fig:nu_profiles}. This effect is not included in the $N$-one-body approach. Furthermore, for the $10^{12}{\rm M}_\odot$ halos it can be seen that the neutrino density falls below its cosmic average beyond the virial radius. This could be due to either the presence of underdensities at particular distances from the halo centers, or due to the fact that we only select isolated halos, which are more likely to be found in low density regions.

From the pure $\Lambda$CDM $N$-body simulations presented in Fig.~\ref{fig:NFW} it can be seen that our matter halos are perfectly fitted by a NFW profile over the mass range $M_{\rm vir}=10^{12}-10^{14} {\rm M}_\odot$ until $20 \, h^{-1} \, {\rm kpc}$ from the halo centers. Here our $N$-body results begin to lack particle resolution. The profile for the larger halo mass is taken from a $1024 \, h^{-1} \, {\rm Mpc}$ box with the same number of particles, and this halo is therefore only resolved until $\sim \, 100 \, h^{-1} {\rm kpc}$. Note that our dominant background NFW profiles in the $N$-body simulation are valid down to scales significantly smaller than the scales at which we present neutrino density profiles. Therefore, our neutrino density profiles are not affected by insufficient CDM $N$-body particle resolution.

Since the CDM component is much more clustered than its neutrino counterpart, the flat profile from the host halo is only dominant relative to the contribution from the halo itself on scales beyond the virial radius (see Fig.~\ref{fig:NFW}). From this figure it can also be readily understood why the neutrino density profiles differ when only low mass isolated halos are considered: The underlying CDM gravitational source term is roughly flat beyond the virial radius, and within the virial radius the neutrinos free-stream out of the small mass halos, in sum producing a roughly flat neutrino density profile also within the virial radius.

\begin{figure}
    \noindent
    \begin{minipage}{1.0\linewidth}
     \begin{minipage}{0.49\linewidth}
           \hspace*{-0.7cm}\includegraphics[width=1.18\linewidth]{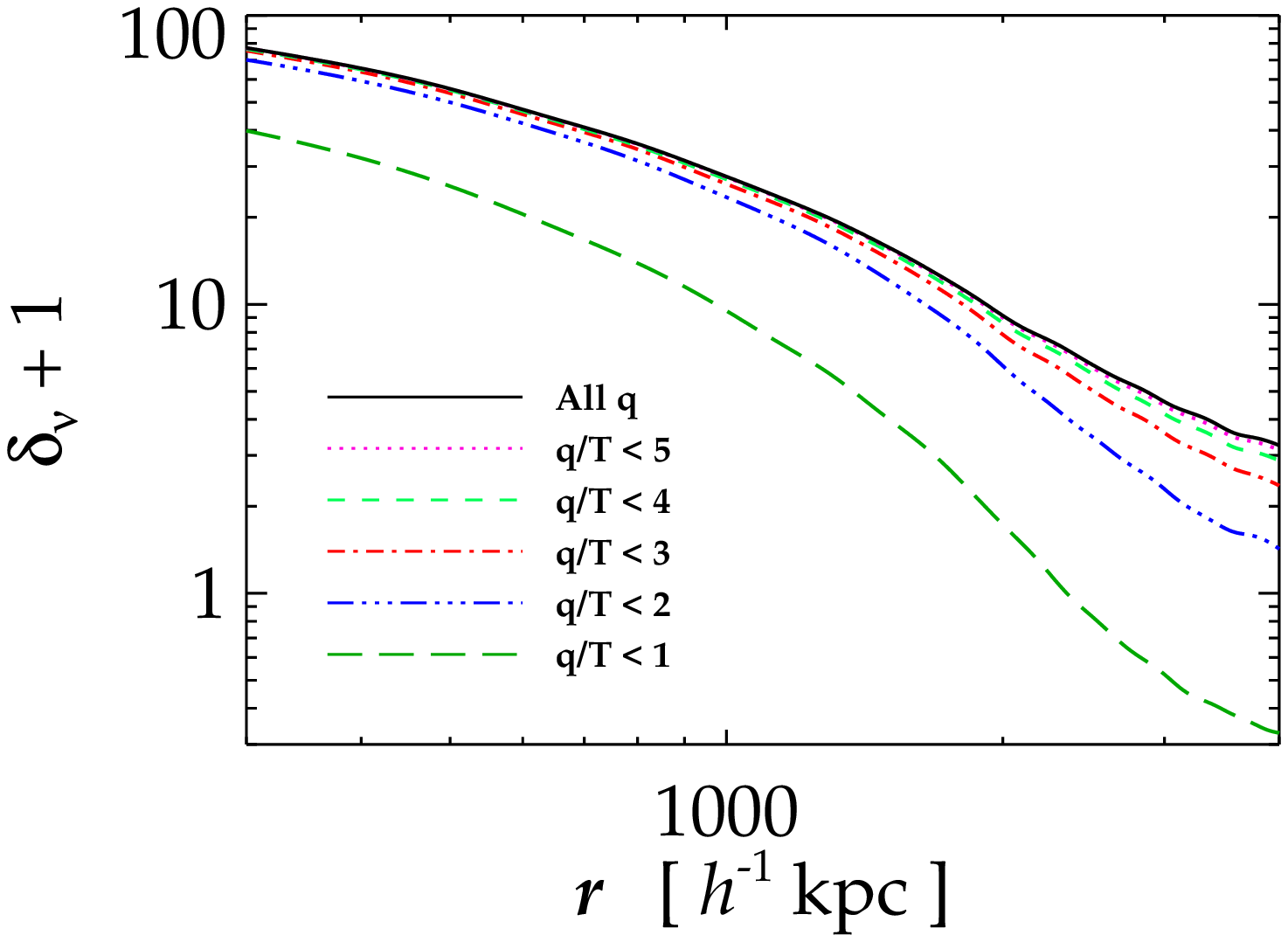}
     \end{minipage}
     \begin{minipage}{0.49\linewidth}
           \hspace*{-0.5cm}\includegraphics[width=1.18\linewidth]{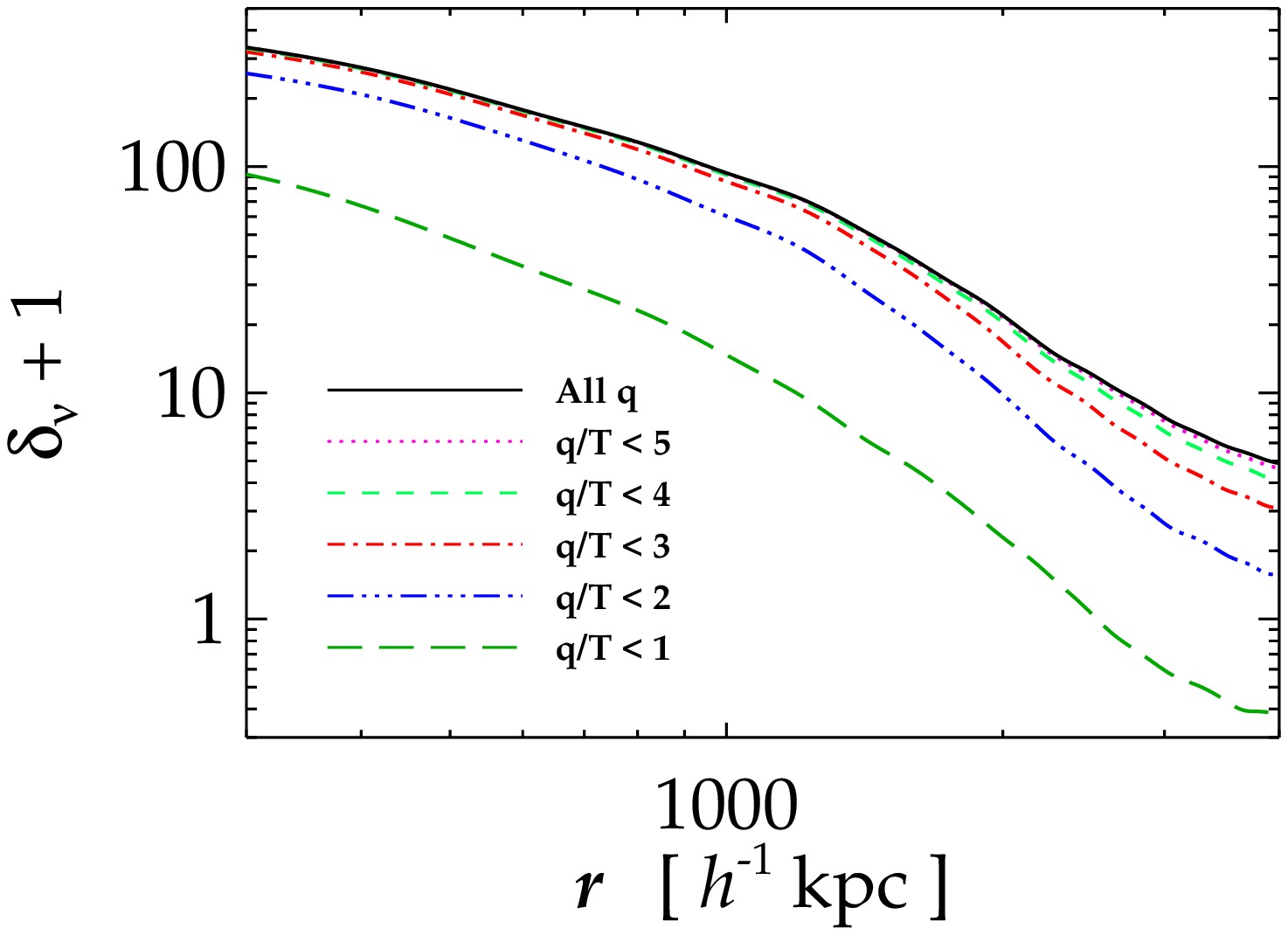}
     \end{minipage}
   \end{minipage}
   \caption{Cumulative neutrino halo density profiles as a function of momentum in a $10^{15}{\rm M}_\odot$ halo, for $\sum m_\nu = 0.6 \, {\rm eV}$ (left) and $\sum m_\nu = 1.2 \, {\rm eV}$ (right). The neutrino density with $q/T > 6$ has been added as a homogeneous term to all profiles.}
   \label{fig:nuprofile2}
\end{figure}

In Fig.~\ref{fig:nuprofile2} we show the cumulative neutrino density profile for a $10^{15}{\rm M}_\odot$ halo, for total neutrino masses of $0.6 \, {\rm eV}$ (left) and $1.2 \, {\rm eV}$ (right). It can be seen that only neutrinos with $q/T < 3$ contribute within the inner $1000 \, h^{-1} \, {\rm kpc}$, whereas neutrinos with momenta up to $q/T \simeq 5$ is needed to simulate profiles beyond $\simeq 2000 \, h^{-1} \, {\rm kpc}$. It can also be seen that as the neutrino mass is increased higher neutrino momentum bins contribute to neutrino clustering on a given scale. Finally, the figures confirm the accuracy of only converting neutrinos with $q/T < 6$ to $N$-body particles. Since this result is accurate for a $10^{15}{\rm M}_\odot$ halo, it is certainly also accurate for lower mass halos where only neutrinos with a momentum from the very low end of the Fermi-Dirac distribution cluster.

\paragraph{The Tremaine-Gunn bound}

Based on purely theoretical grounds one should expect the following scenario. Neutrinos cluster in a halo of given mass $M_{\rm vir}$ and radius $r_{\rm vir}$, so that the escape velocity at $r_{\rm vir}$ of a given halo is
\begin{equation}
v_e \sim \sqrt{\frac{M_{\rm vir}}{r_{\rm vir}}}.
\end{equation}
Only neutrinos up to this velocity can be bound in the halo and the central density should therefore be
\begin{equation}
\rho_\nu \sim \int_0^{v_e m_\nu} m_\nu p^2 f(p) dp.
\end{equation}
For small $p$, $f$ is approximately $1/2$, and we have
\begin{equation}
\rho_\nu \sim \left(\frac{M_{\rm vir}}{r_{\rm vir}} \right)^{3/2} m_\nu^4,
\end{equation}
leading to
\begin{equation}
\delta_\nu \sim \left(\frac{M_{\rm vir}}{r_{\rm vir}} \right)^{3/2} m_\nu^3.
\end{equation}
This is the Tremaine-Gunn (TG) bound~\cite{Tremaine:1979we}, and is essentially the same result quoted in Eq.~(7.3) of \cite{Ringwald:2004np} for a NFW halo.

\begin{figure}
  \noindent
    \begin{minipage}{1.0\linewidth}
     \begin{minipage}{0.49\linewidth}
           \hspace*{-0.7cm}\includegraphics[width=1.18\linewidth]{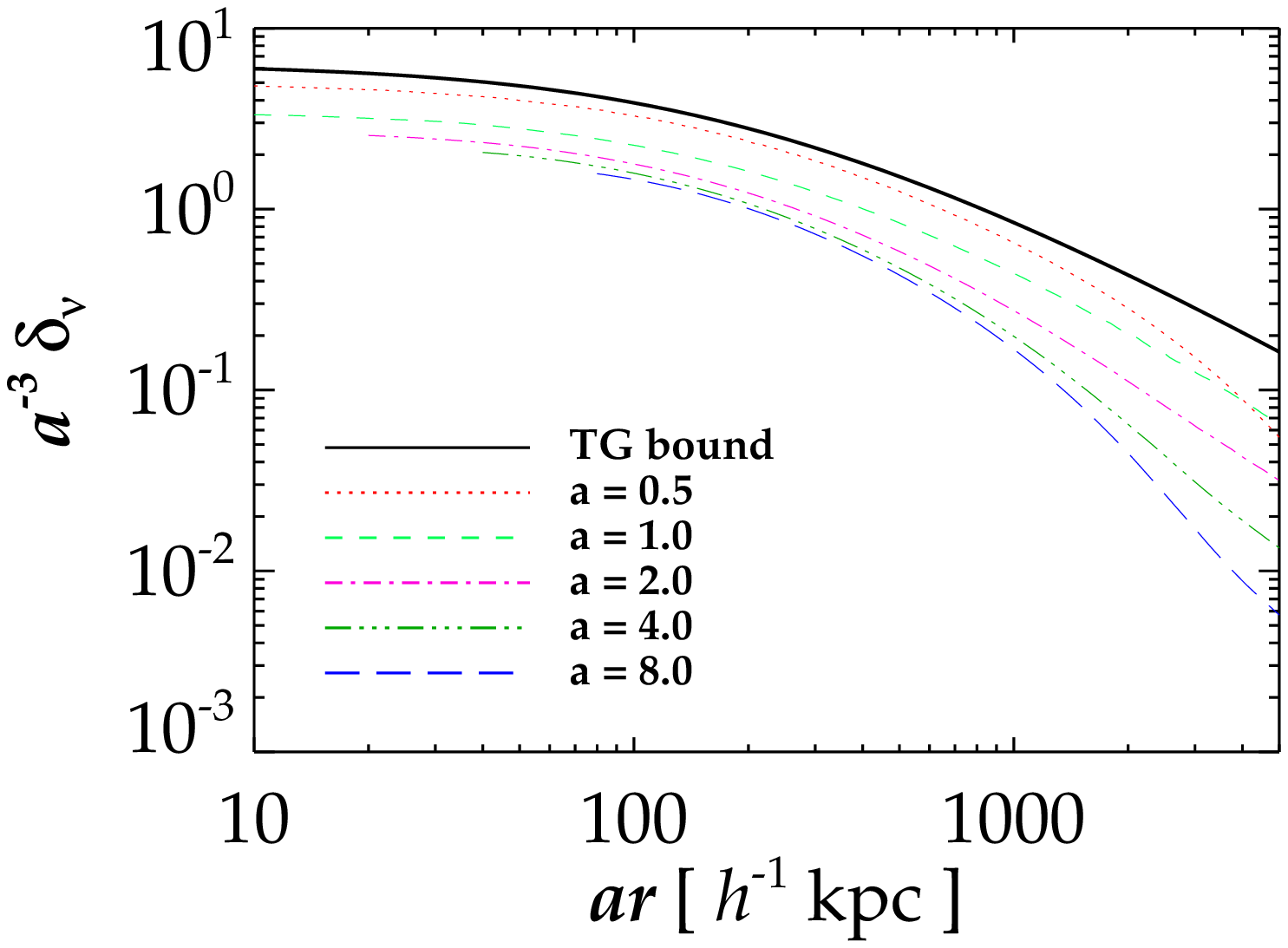}
     \end{minipage}
     \begin{minipage}{0.49\linewidth}
           \hspace*{-0.5cm}\includegraphics[width=1.18\linewidth]{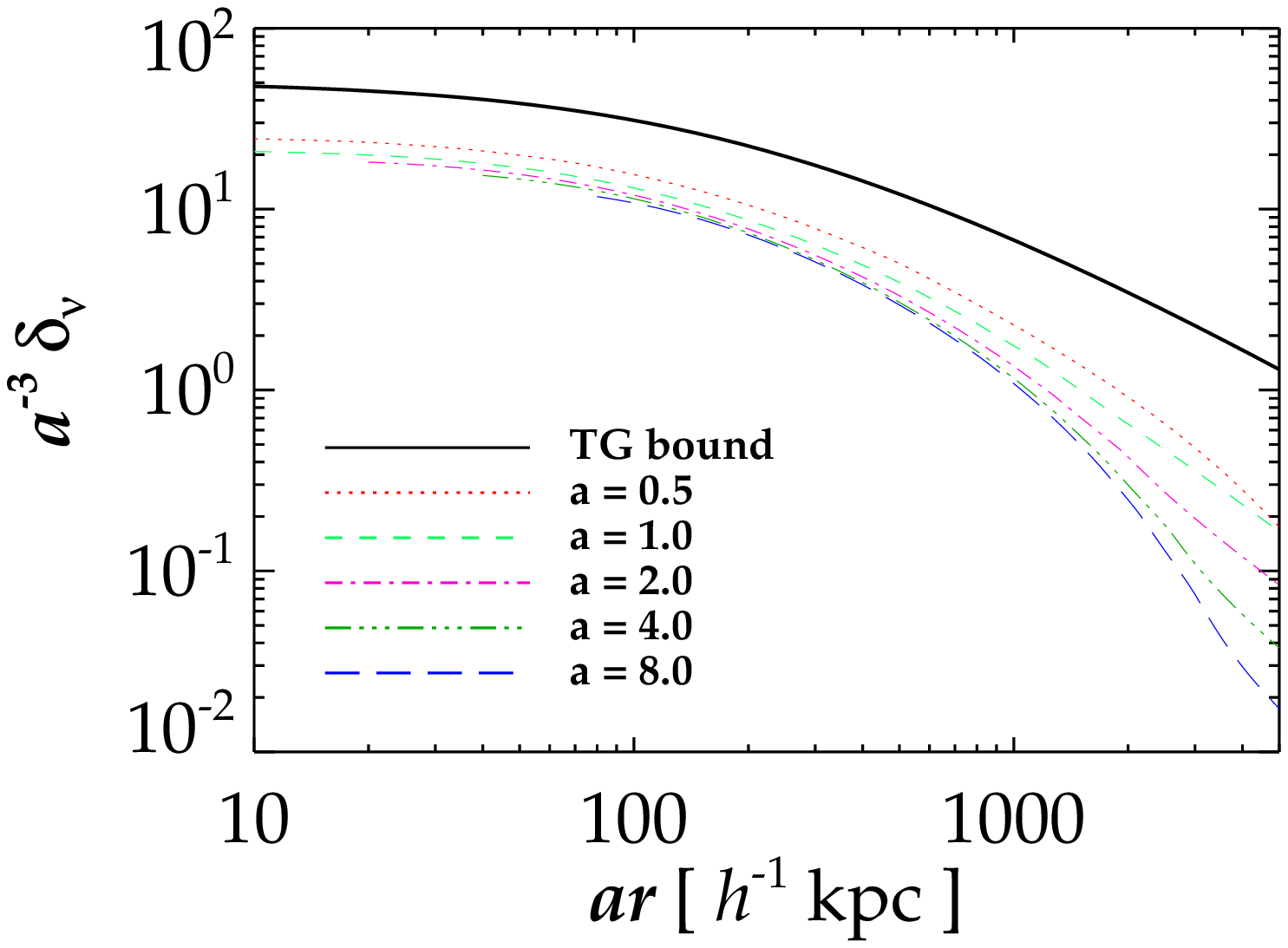}
     \end{minipage}
   \end{minipage}
   \caption{The evolution of the neutrino density profile around a static NFW halo of $10^{14} {\rm M}_\odot$ for $\sum m_\nu = 0.3$ eV (left) and  $\sum m_\nu = 0.6$ eV (right). The physical density perturbation, $a^{-3}\delta_\nu$, as a function of the physical radius, $ar$, is time-independent in the TG limit.}
   \label{fig:tg1}
\end{figure}

However, in practice the neutrino halo density almost never saturates the Tremaine-Gunn bound, as was also seen in Fig.~7 of \cite{Ringwald:2004np}.
Fig.~\ref{fig:tg1} sheds some light on the actual evolution of the neutrino halos. The setting is a simplified version of the $N$-one-body method in which the NFW halo is taken to be {\it static} so that the TG bound is time-independent in physical units. For the smallest neutrino mass $\sum m_\nu = 0.3$ eV, it can be seen that the physical density perturbations of neutrinos drop as the scale factor increases, stabilising only at $a \sim 4$. At early times, i.e., $a=1/2$, the TG bound is almost saturated because the neutrino density contrast is low and
the halo is populated only with neutrinos drawn from the very low momentum end of the relativistic Fermi-Dirac distribution for which $f \sim 1/2$.  However, this also means that the final neutrino phase
space density deviates little from the relativistic Fermi-Dirac distribution, so that linear perturbation theory remains valid. The dilution due to the background expansion dominates until very late and the physical neutrino density perturbation only approaches a constant at $a \sim 4$, and at a value much lower than the TG bound.

For the higher neutrino mass this effect is less pronounced, i.e., the physical density levels off much earlier. However, for the higher neutrino mass the TG bound is far from saturated even at early times. The reason for this can be understood from Fig.~\ref{fig:nuprofile2}: For the higher neutrino mass, neutrinos with $q/T > 1$ make up most of the halo.  However, since the initial neutrino phase space density is far less than $1/2$ at $q/T > 1$, it also makes
it more difficult for the final coarse-grained phase space density to reach the upper limit of $f \sim 1/2$ at the higher momentum end of the spectrum.
For example, the second bin with $1 < q/T < 2$ has $\bar{f} = \int_1^2 f^2(q) q^2 dq/ \int_1^2 f(q) q^2 dq \sim 0.18$.

\subsection{Feedback on CDM halos}

\begin{figure}
    \noindent
    \begin{minipage}{1.0\linewidth}
     \begin{minipage}{0.49\linewidth}
           \hspace*{-0.7cm}\includegraphics[width=1.18\linewidth]{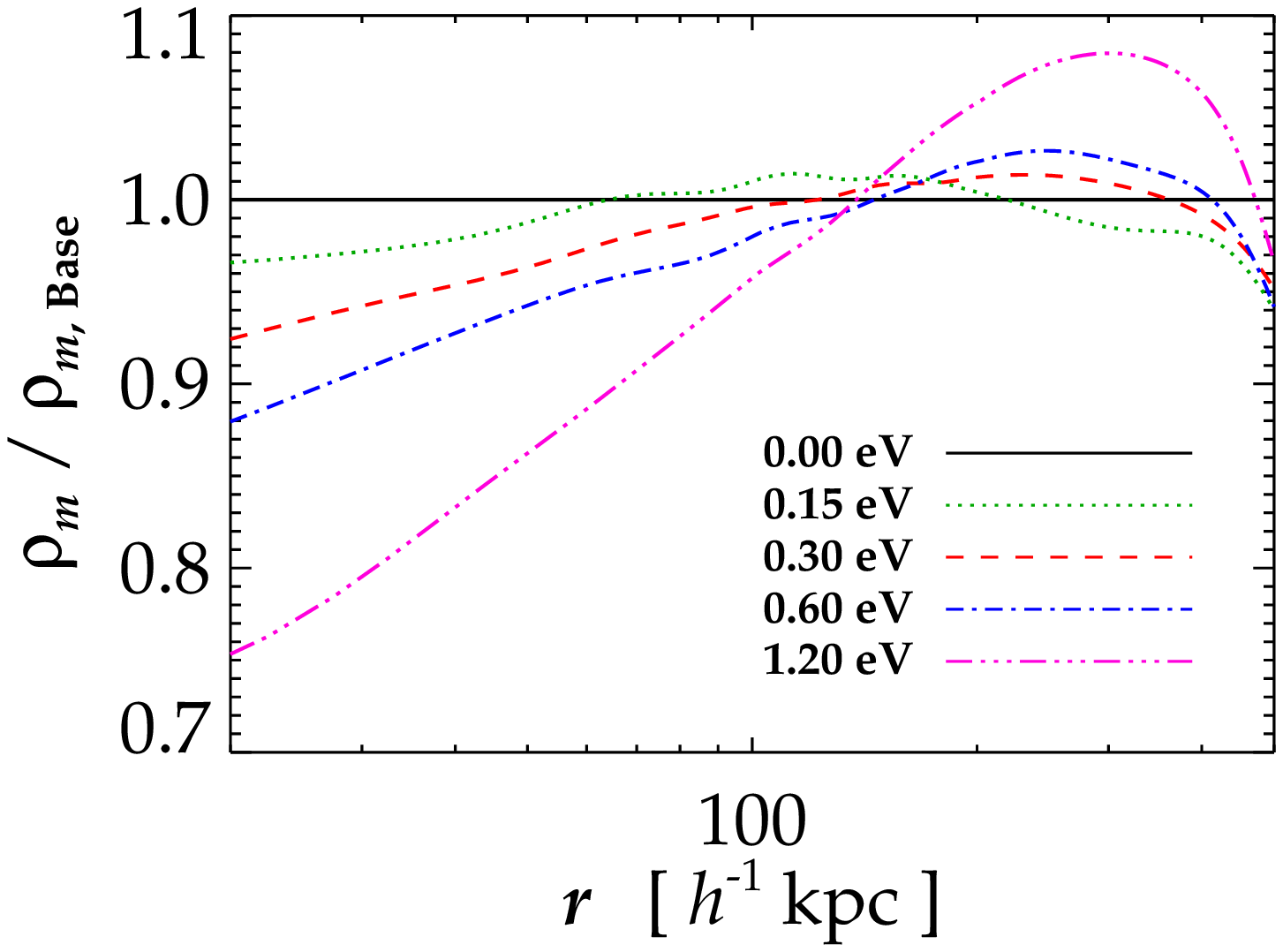}
     \end{minipage}
     \begin{minipage}{0.49\linewidth}
           \hspace*{-0.5cm}\includegraphics[width=1.18\linewidth]{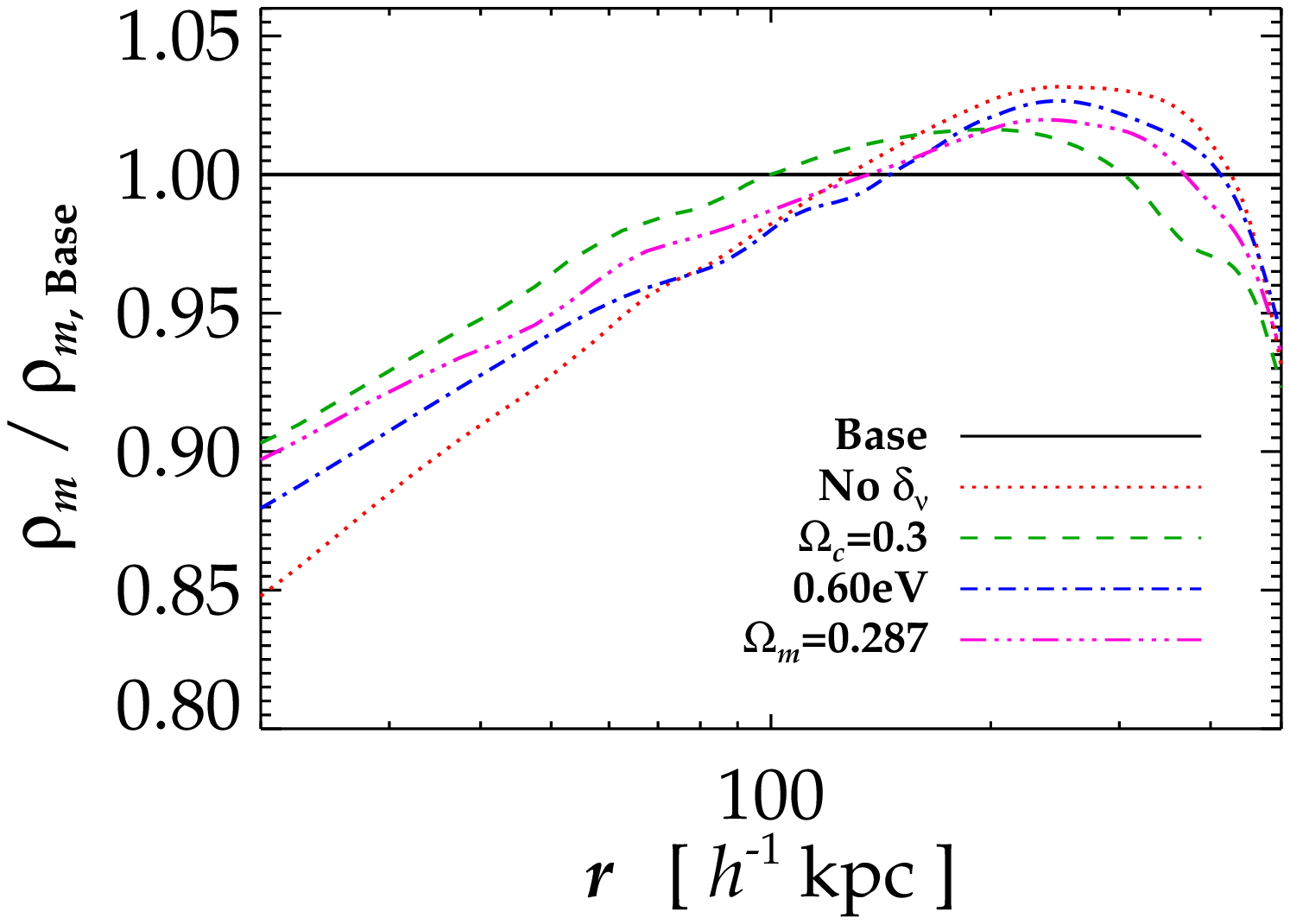}
     \end{minipage}
   \end{minipage}
    \caption{Change in the matter halo profiles relative to a base model without massive neutrinos for a halo mass of $10^{13} \, {\rm M}_\odot$. Left: As a function of neutrino mass. Right: For different (exotic) cosmologies (see details in text).}
   \label{fig:relative_profiles}
\end{figure}

Even though neutrinos make up a minute fraction of the dark matter halo mass at late times, they do have an impact on the halo formation history, i.e., they affect the halo mass dependent merger rate \cite{Fakhouri:2010st}. In this section we are interested in alterations of the halo profiles, not the number density of halos in a given mass interval (i.e., the halo mass function). This aspect will be discussed in detail in Section \ref{sec:massfunction}.

Fig.~\ref{fig:relative_profiles} (left) shows the relative change of the matter halo density profile as a function of neutrino mass for a $10^{13} \, {\rm M}_\odot$ halo. We only show results for this halo mass, since here the product of the number density of halos times the halo density is maximal. From the figure it can be seen that the presence of massive neutrinos lowers the density at radii smaller than $\sim 100 \, h^{-1} \, {\rm kpc}$. Since we compare halos of identical total masses this is compensated by an increased density at larger radii. The reason for this effect is that halos form later in models with massive neutrinos because the linear theory TF is lowered. From pure CDM simulations it is indeed known that late forming halos are less concentrated \cite{Wechsler:2001cs}, with the concentration parameter scaling roughly as $c \propto 1/a_c$, where $a_c$ is the scale factor at formation.

From the right hand side of Fig.~\ref{fig:relative_profiles} it can be seen that removing the neutrino perturbations (labelled $\delta_\nu = 0$) from a cosmology with $\sum m_\nu = 0.6 \, {\rm eV}$ in the $N$-body simulation leads to less concentrated halos. On the scales shown the suppresion is as large as $3-4\%$, though we caution that there is some noise in the data. It can also be seen that replacing $\Omega_\nu$ by an enlarged $\Omega_c$ (labelled $\Omega_c = 0.3$) when the $N$-body simulation is started gives a profile which is too steep at the few \% level. Finally, a cosmology with $\Omega_m=0.287$ (labelled $\Omega_m = 0.287$) is placed somewhere between the aforementioned cosmologies.

%%%%%%%%%%%%%%%%%%%%%%%%%%%%%%%%%%%%%%%%%%%%%%%%%%%%%%%%%%%%%%%%%%%%%%%%

\section{The halo mass function}
\label{sec:massfunction}

\begin{figure}
    \noindent
     \begin{center}
           \includegraphics[width=0.7\linewidth]{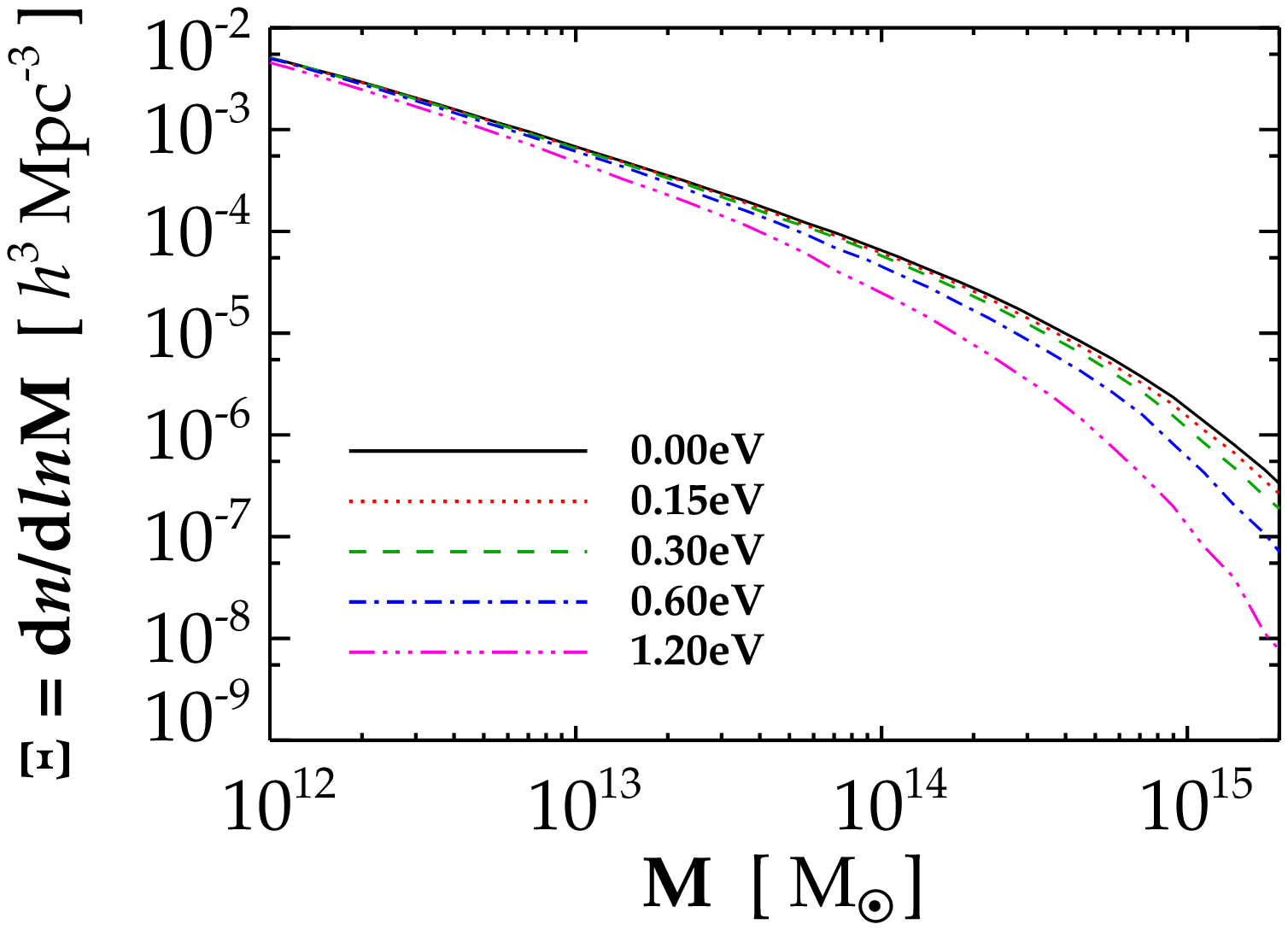}
     \end{center}
     \vspace*{-1.2cm}

     \begin{center}
           \includegraphics[width=0.7\linewidth]{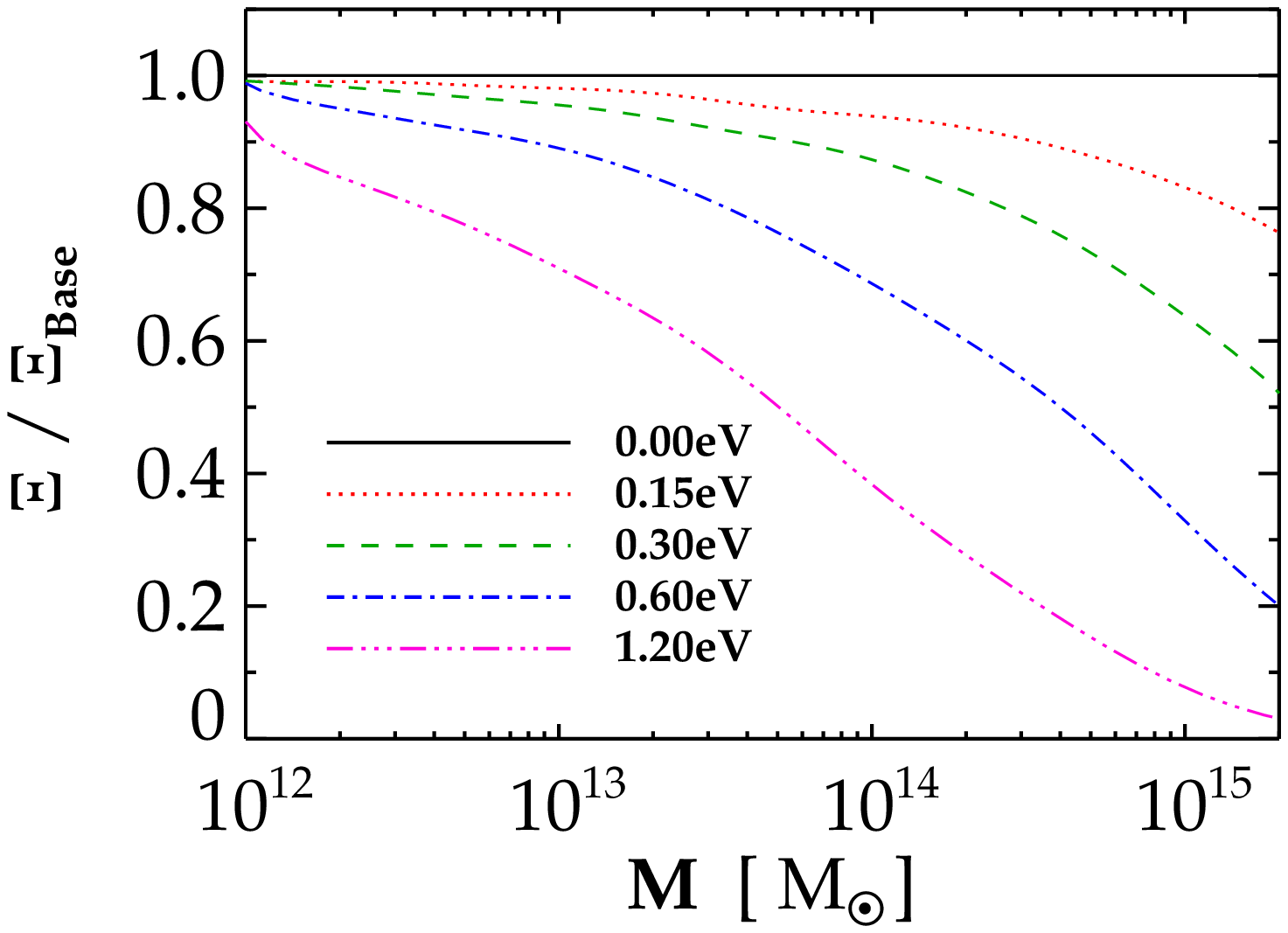}
     \end{center}
     \vspace*{-1.2cm}

     \begin{center}
           \includegraphics[width=0.7\linewidth]{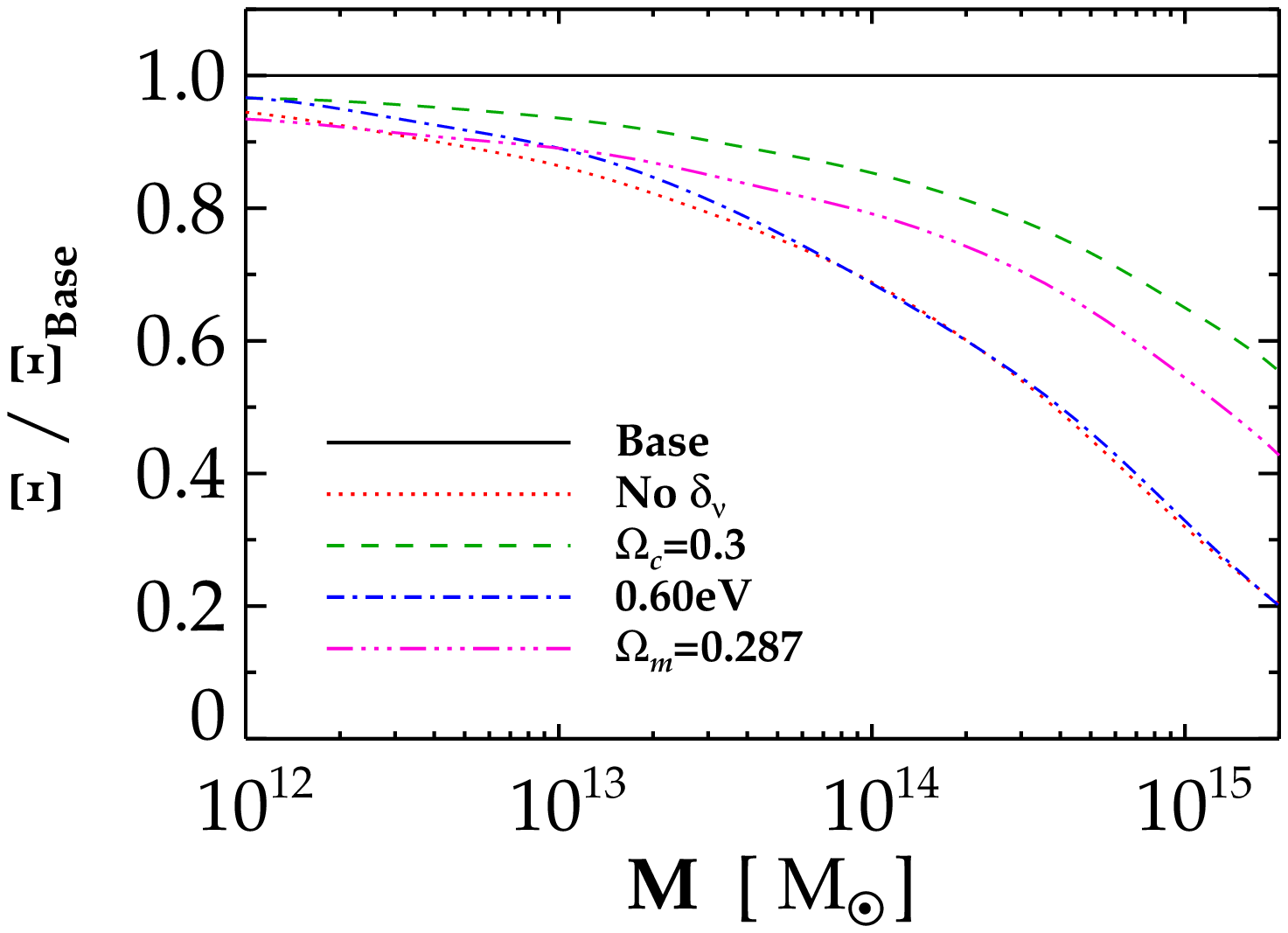}
     \end{center}
     \vspace*{-0.5cm}

   \caption{Absolute (top) and relative (middle) halo mass functions for 5 different neutrino cosmologies. The halo mass functions have been splined and smoothed together to obtain sufficient accuracy in the halo mass range $10^{12}$ to $10^{15}$ ${\rm M}_\odot$. Bottom: Relative change in our halo mass function for different (exotic) cosmologies.}
   \label{fig:halo_mass_function}
\end{figure}

\begin{figure}
    \noindent
     \begin{center}
           \includegraphics[width=0.7\linewidth]{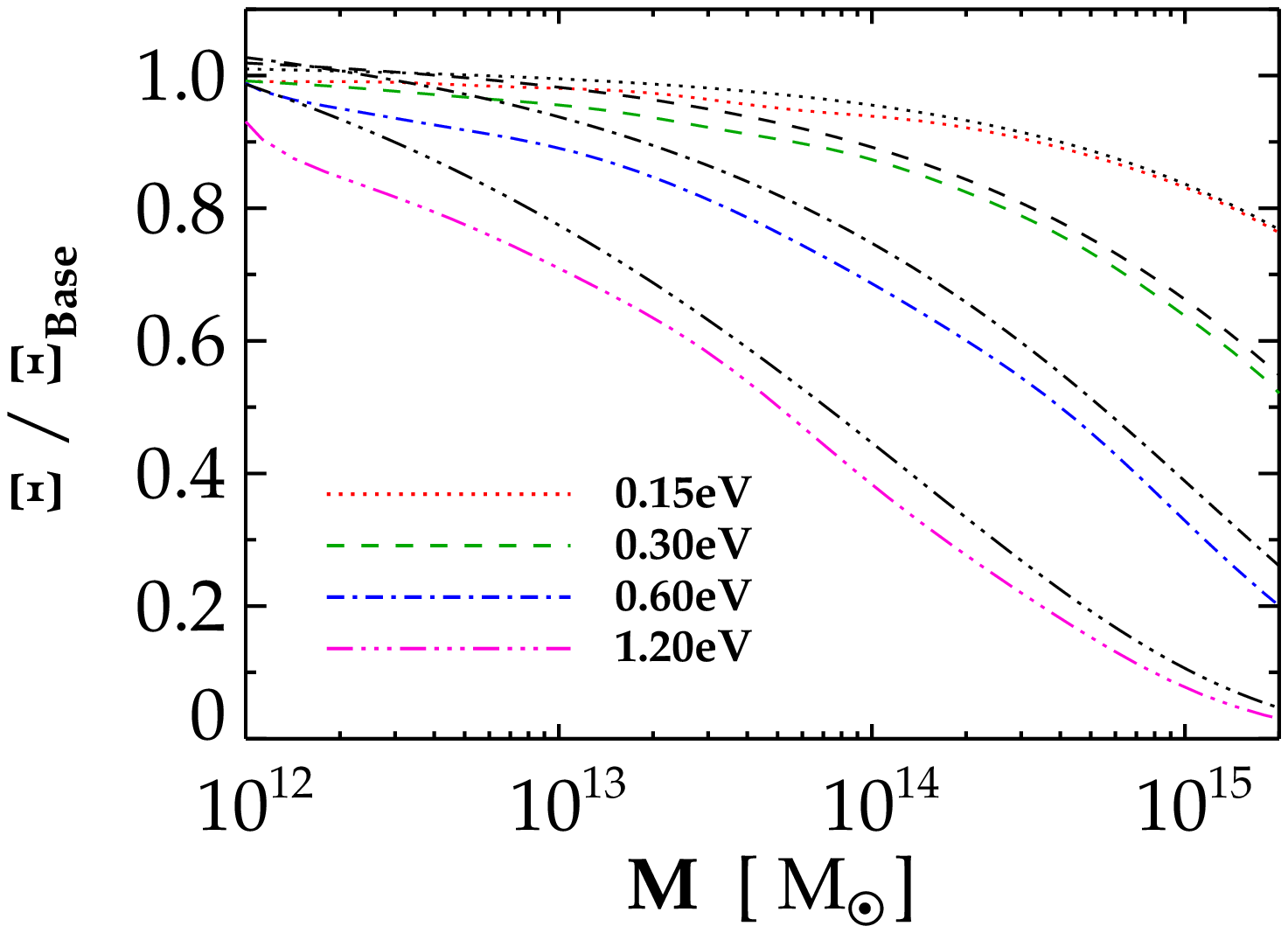}
     \end{center}
     \vspace*{-1.2cm}

     \begin{center}
           \includegraphics[width=0.7\linewidth]{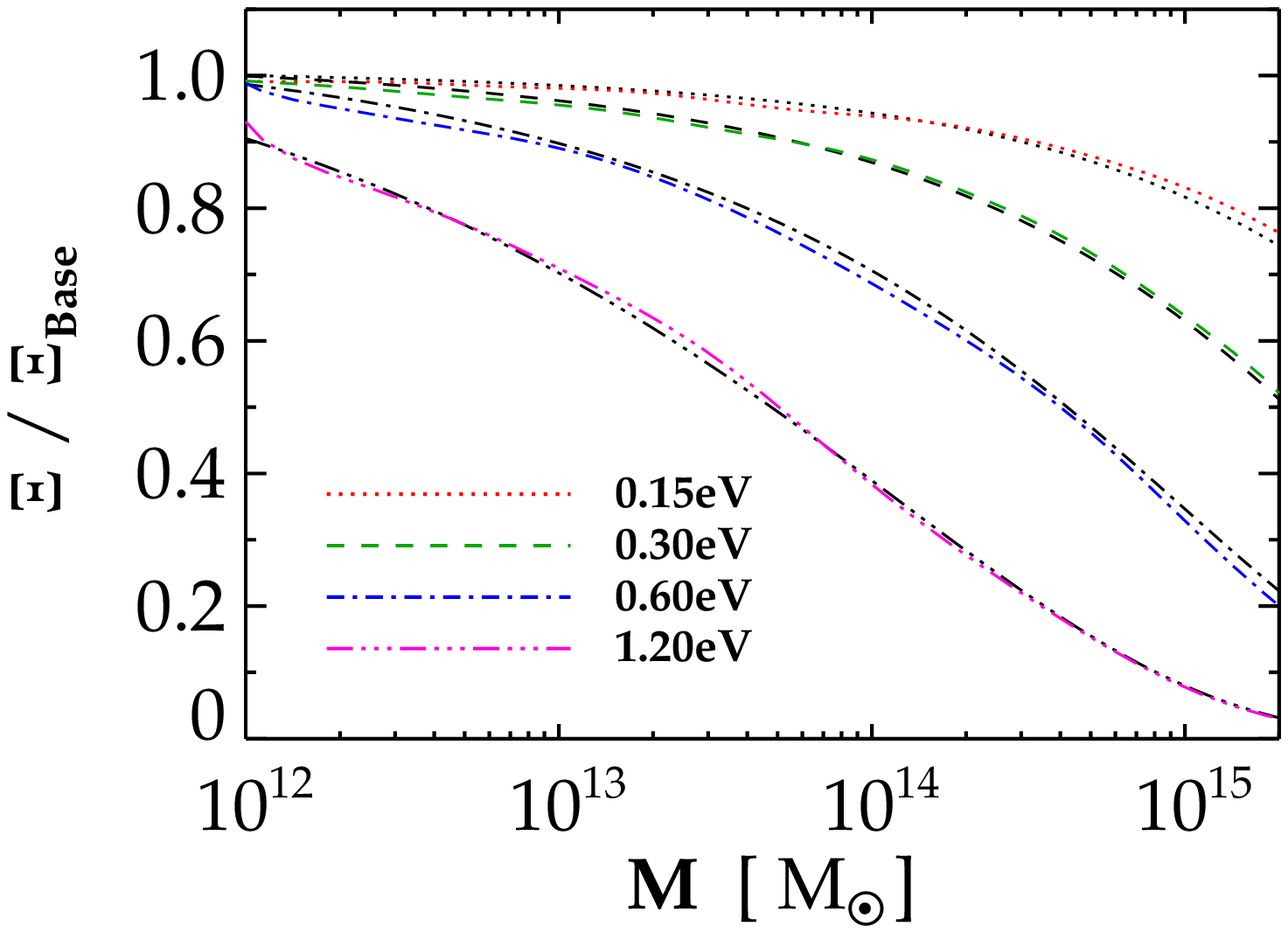}
     \end{center}
     \vspace*{-1.2cm}

     \begin{center}
           \includegraphics[width=0.7\linewidth]{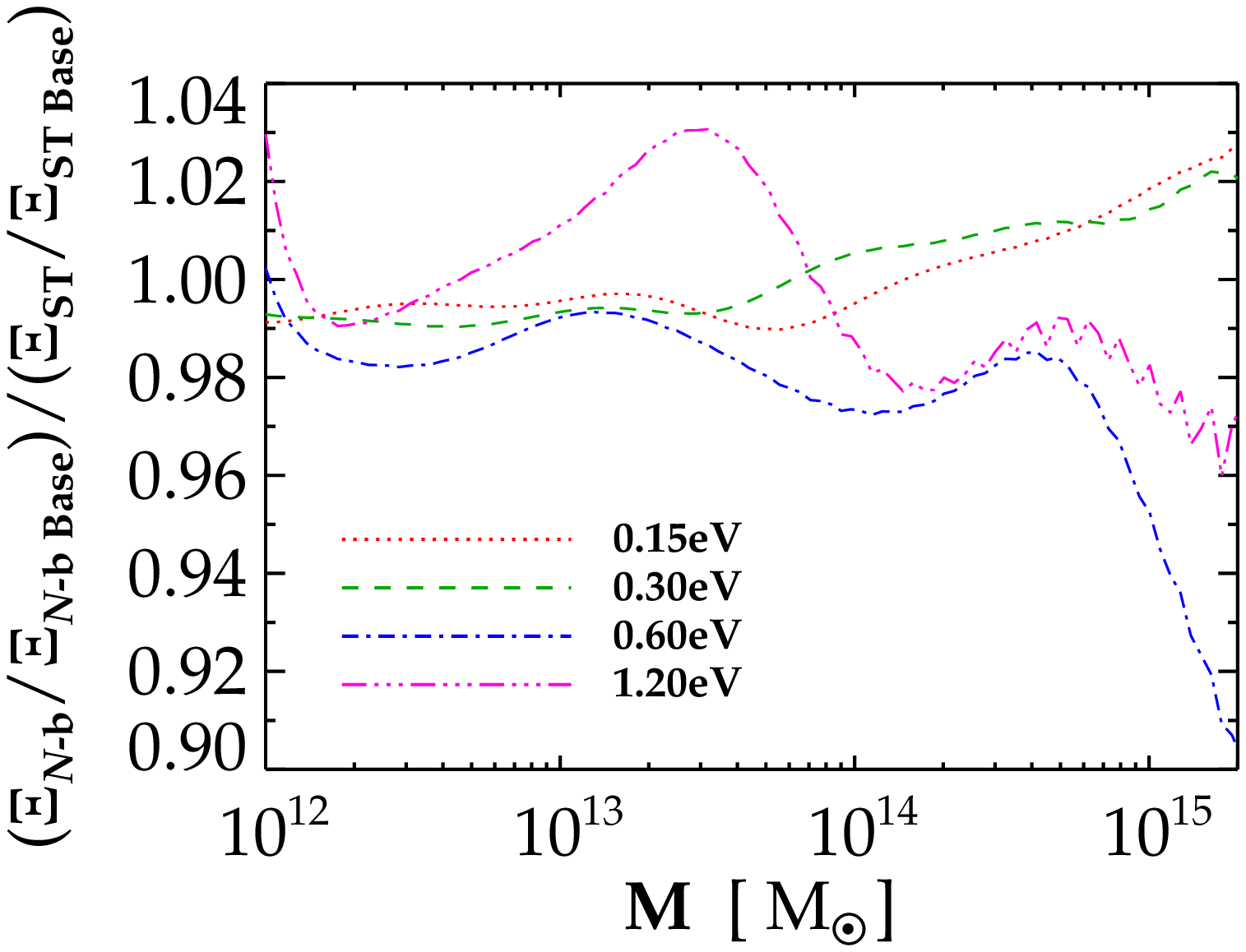}
     \end{center}
     \vspace*{-0.5cm}

   \caption{Relative halo mass functions for different neutrino cosmologies compared with the predictions from the Sheth-Tormen formulae (black lines). Top: With $\Omega_m = \Omega_c + \Omega_b + \Omega_\nu$ in the ST formulae. Middle: With $\Omega_c + \Omega_b$ used instead of $\Omega_m$ in the ST formulae. Bottom: Differences between the $N$-body and the ST predictions.}
   \label{fig:shethtormen}
\end{figure}

We now turn our attention to the number density of halos in a given mass interval, the halo mass function (HMF). In Fig.~\ref{fig:halo_mass_function} (top and middle figures) we show the HMFs for cosmologies with different neutrino masses. As expected the HMF is more suppressed in cosmologies with a larger neutrino mass. The suppression is largest for the heaviest, late forming halos.

The last panel in Fig.~\ref{fig:halo_mass_function} shows that the cosmologies where the $\sum m_\nu = 0.6 \, {\rm eV}$ neutrino component is replaced by CDM at the redshift where the $N$-body simulation is started as well as the one with $\Omega_m = 0.287$ predict HMFs with different shapes. Only a model without neutrino perturbations in the $N$-body simulation gives roughly the same result as the full calculation:
The suppression of the HMF is mainly caused by the suppression of the initial TF in the linear regime and not by neutrino clustering effects in the $N$-body simulation.

In Fig.~\ref{fig:shethtormen} we compare our HMFs calculated from $N$-body simulations with HMFs from the Sheth-Tormen (ST) semi-analytic formulae \cite{Sheth:2001dp}.\footnote{See \cite{Jenkins:2000bv,Warren:2005ey} for other semi-analytic fits to the mass function, which we found to work less well for models with massive neutrinos.}
The ST fit is based on the fact that, as first pointed out by Press and Schechter \cite{Press:1973iz}, the HMF can be written as
\begin{equation}
\frac{M dM}{\bar \rho}\frac{dn(M,z)}{dM} = \nu f(\nu) \frac{d\nu}{\nu},
\end{equation}
with $\nu \equiv [\delta_{\rm sc}(z)/\sigma(M)]^2$, where $\delta_{\rm sc}(z)=1.686$ is the overdensity required for spherical collapse at $z$, and $\bar \rho = \Omega_m \rho_c$.
$dn(M,z)$ is the number density of halos in the mass interval $M$ to $M + dM$.
The variance of the linear theory density field, $\sigma^2(M)$, is given by
\begin{equation}
\sigma^2(M) = \int \frac{dk}{k} \frac{k^2 P_{\rm lin}(k)}{2 \pi^2} |W(kR)|^2,
\end{equation}
where $P_{\rm lin}(k)$ is the linear theory matter power spectrum, and the Top-Hat window function is given by $W(x) = (3/x^2)(\sin x - x \cos x)$ with $R = (3M/4\pi \bar \rho)^{1/3}$.

The ST fit to $\nu f(\nu)$ is
\begin{equation}
\nu f(\nu) = A \left(1+\frac{1}{\nu'^p}\right)\left(\frac{\nu'}{2}\right)^{1/2} \frac{e^{-\nu'/2}}{\sqrt{\pi}},
\end{equation}
with $\nu' = 0.707 \nu$ and $p=0.3$. $A=0.3222$ is determined from the integral constraint $\int f(\nu) d\nu = 1$.

The upper panel in Fig.~\ref{fig:shethtormen} shows that the agreement is poor if $\Omega_m = \Omega_c + \Omega_b + \Omega_\nu$ is used in the ST formalism. However, this is due to a wrong definition of the halo mass: Even for the very largest cluster halos the neutrino component contributes very little to the halo mass. In reality, the mass inside the collapsing region should be calculated using $\Omega_c + \Omega_b$, not $\Omega_m$. This amounts to neglecting the weakly clustering neutrino component when calculating the halo mass.
The two lower panels in Fig.~\ref{fig:shethtormen} shows the same ST fit, but using $\Omega_c + \Omega_b$ instead of $\Omega_m$. In this case the ST HMFs provide an excellent fit to the
relative change to the HMF caused by neutrinos. As the figure at the bottom clearly demonstrates, the agreement is better than $\sim 3\%$ at halo mass scales where our $N$-body HMFs are accurate (we do not consider our $N$-body HMFs to be accurate by more than a few \% on any mass scale, $M$).
Although the absolute HMFs, even for CDM simulations, do not match the ST HMFs more precisely than at the $\sim 10$\% level, the {\it relative} change from adding neutrinos can be calculated significantly more accurately.

%%%%%%%%%%%%%%%%%%%%%%%%%%%%%%%%%%%%%%%%%%%%%%%%%%%%%%%%%%%%%%%%%%%%%%%%%%%%%%%%%%%%%%%%%%%%%%%%%%%%%%%%%%%%%%%

\section{Conclusion}\label{sec:conclusion}

We have performed a detailed study of halo properties in $\Lambda$CDM cosmologies with massive neutrinos included. An important goal was to study the neutrino density profiles in dark matter halos. To this end we employed detailed $N$-body simulations across a wide range of scales to test halo masses from Milky Way size ($10^{12} \, {\rm M}_\odot$) to large clusters ($10^{15} \, {\rm M}_\odot$), as well as the $N$-one-body method developed to solve the neutrino Boltzmann equation approximately around existing CDM halos. In general we found good agreement between the full $N$-body and the $N$-one-body results.
The difference between the $N$-body and $N$-one-body methods arise from the fact that the latter assumes the CDM halo to be monolithic and at all times describable in terms of a NFW profile, i.e., it does not take into account halo substructure and larger merger events. It also assumes an analytic evolution of the concentration parameter.
We also discussed in some detail how the density profiles of neutrino halos can be understood in terms of the Tremaine-Gunn bound, i.e., the bound coming from the fact that a coarse-grained distribution can never attain values exceeding the maximum of the original fine-grained distribution.

For smaller halo masses, the neutrino profiles in isolated halos are in excellent agreement with the prediction from the $N$-one-body method. This result is not too surprising since this is exactly the case where the infall on an existing spherical NFW halo is most realistic. However, many smaller mass halos are embedded in larger cluster halos and for the smaller neutrino masses the local neutrino profile in such a halo is dominated by the background of neutrinos bound in the much larger cluster halos.

In terms of the local neutrino density enhancement, which is relevant for possible future attempts at direct C$\nu$B detection, a Milky Way-size galaxy halo is too small to have a significant overdensity, even when taking a possible cluster background into account.

We also briefly studied how neutrinos impact on the density profiles of the CDM halos. While neutrinos contribute very little to the total density in the halo, the presence of massive neutrinos in the model leads to slightly later formation of halos with a given mass and consequently to generally lower concentration parameters, $c$.

Finally, we calculated halo mass functions for $\Lambda$CDM models with massive neutrinos. Since large cluster surveys will become available in the coming years, the halo mass function is an important cosmological observable.
As expected, we find a very strong suppression of halo formation with increasing neutrino mass. As noted in previous analytic or semi-analytic studies the suppression is particularly marked for massive halos because the suppression in linear theory power from massive neutrinos shifts the maximum cluster mass down, i.e.\ the scale beyond which the halo mass function is exponentially suppressed.

We then compared the halo mass functions from simulations with halo mass functions calculated using the semi-analytic method developed by Sheth and Tormen. If used na\"{\i}vely, i.e.\ just processing the linear theory power spectrum without any adjustment to the method, the agreement is poor. However, it is easy to see that the disagreement arises because the ST method implicitly assumes that all matter clusters in the same way (the value of $\Omega_m$ used is $\Omega_c+\Omega_b+\Omega_\nu$). However, even large clusters bind relatively few neutrinos and for all halos it is true that neutrinos make a negligible contribution to the halo mass. If the ST formalism is corrected for this by using $\Omega_m=\Omega_c+\Omega_b$, i.e.\ taking into account only the clustering species (but of course using the correct initial power spectrum and the correct background evolution), the agreement between the modified ST and the $N$-body results is remarkable. On all measurable scales it is better than 2-3\%. This is important for analysing future cluster surveys because it means that existing semi-analytic methods can be used instead of having to perform time consuming simulations for all neutrino masses.

Alternatively, neglecting the neutrino perturbations in the $N$-body simulation will also be a very accurate approximation for $\sum m_\nu \lesssim 0.5 \, {\rm eV}$ as long as only matter halo properties are considered. This approximation is not valid for a precise calculation of the matter power spectrum \cite{Brandbyge1}. In general the accuracy of the approximation is determined by contrasting the neutrino free-streaming length with the physical extent of the scales simulated: Considering halo properties and realistic neutrino masses, this approximation is very good.

%%%%%%%%%%%%%%%%%%%%%%%%%%%%%%%%%%%%%%%%%%%%%%%%%%%%%%%%%%%%%%%%%%%%%%%%%%%%%%%%%%%%%%%%%%%%%%%%%%%%%%%%%%%%%%%%%%%%%%%%%%%%%%%%%%%%%%
%%%%%%%%%%%%%%%%%%%%%%%%%%%%%%%%%%%%%%%%%%%%%%%%%%%%%%%%%%%%%%%%%%%%%%%%%%%%%%%%%%%%%%%%%%%%%%%%%%%%%%%%%%%%%%%%%%%%%%%%%%%%%%%%%%%%%%
\section*{Acknowledgements}
We acknowledge computing resources from the Danish Center for
Scientific Computing (DCSC).

%%%%%%%%%%%%%%%%%%%%%%%%%%%%%%%%%%%%%%%%%%%%%%%%%%%%%%%%%%%%%%%%%%%%%%%%%%%%%%%%%%%%%%%%%%%%%%%%%%%%%%%%%%%%%%%%%%%%%%%%%%%%%%
%%%%%%%%%%%%%%%%%%%%%%%%%%%%%%%%%%%%%%%%%%%%%%%%%%%%%%%%%%%%%%%%%%%%%%
%%%%%%%%%%%%%%%%%%%%%%%%%%%%%%%%%%%%%%%%%%%%
\section*{References} %%%%%%%%%%%%%%%%%%%%%%%%%%%%%%%%%%%%%%%%%%%%%%%%
%%%%%%%%%%%%%%%%%%%%%%%%%%%%%%%%%%%%%%%%%%%%%%%%%%%%%%%%%%%%%%%%%%%%%%

\end{document}